\documentclass[aps, pra,reprint, showpacs,nofootinbib,
superscriptaddress,10pt,floatfix,longbibliography]{revtex4-2}
\usepackage[utf8]{inputenc}
\usepackage{amsmath, amssymb, braket, mathtools, amsthm}
\usepackage[qm]{qcircuit}
\usepackage{graphicx}
\usepackage{ragged2e}
\usepackage[ruled,vlined]{algorithm2e}
\usepackage{bm}
\usepackage[normalem]{ulem}
\usepackage{physics}
\usepackage{hyperref}
\usepackage{orcidlink}
\usepackage[caption=false]{subfig} 
\usepackage{xcolor}     
\usepackage{microtype}         
\usepackage{natbib}
\usepackage{ulem}
\usepackage{soul}
\usepackage{etoolbox}    
\AtBeginEnvironment{thebibliography}{%
  \sloppy
  \setlength\emergencystretch{1em}%
}
\AtEndEnvironment{thebibliography}{\fussy}

\allowdisplaybreaks
\SetKw{KwStep}{step}
\definecolor{deepGreen}{HTML}{057a20}
\definecolor{rkrPurple}{HTML}{73024F}
\definecolor{rkrblue}{HTML}{13b2ba}

\newtheorem{theorem}{\bf Theorem}
\newtheorem{proposition}{\bf Proposition}
\newtheorem{definition}{\bf Definition}
\newtheorem{lemma}{\bf Lemma}
\newtheorem{corollary}{\bf Corollary}
\newcommand{\supt}[1]{\bm{\tilde{#1}}}
\newcommand{\cnsub}[2]{\textsc{CNOT}$_{(#1,\,#2)}$}
\newcommand{\cnsubM}[2]{\textsc{CNOT}_{(#1,\,#2)}}
\newcommand{\cn}{\textsc{CNOT}}
\renewcommand{\th}{\text{th}}
\newcommand{\curly}[1]{\{#1\}}
\newcommand{\bigcurly}[1]{\Big\{#1 \Big\}}
\newcommand\struct{\stackrel{\mathclap{\normalfont\mbox{\tiny SS}}}{=}}

\newcommand{\pcsadd}{Center for Theoretical Physics of Complex Systems, Institute for Basic Science(IBS), Daejeon 34126, Republic of Korea}

\begin{document}


\title{Quantum circuit model for continuous-time quantum walks on random graphs}

\author{Sabyasachi Chakraborty\orcidlink{0009-0001-8084-0565}}
\email{sabyasachi.sch@gmail.com}
\affiliation{Department of Physics, Indian Institute of Technology Kharagpur, Kharagpur, West Bengal 721302, India}

\author{Rohit Sarma Sarkar\orcidlink{0000-0003-0747-2572}}
\email{rohit15sarkar@yahoo.com}
\affiliation{International Centre for Theoretical Sciences (ICTS-TIFR), Bengaluru 560089, India}

\author{Sonjoy Majumder\orcidlink{0000-0001-9131-4520}}
\email{sonjoym@phy.iitkgp.ac.in}
\affiliation{Department of Physics, Indian Institute of Technology Kharagpur, Kharagpur, West Bengal 721302, India}

\author{Rohit Kishan Ray\orcidlink{0000-0002-5443-4782}}
\email{rkray@vt.edu}
\affiliation{Department of Material Science and Engineering, Virginia Tech, Blacksburg, VA 24061, USA}
\address{\pcsadd}


\begin{abstract}
Quantum-circuit implementations of continuous-time quantum walks (CTQWs) can provide an efficient route to model graph-based algorithms. However, constructing circuits that faithfully reproduce CTQW dynamics across arbitrary graphs remains a major challenge. In this work, we introduce a Laplacian partitioning algorithm (LPA) that enables an efficient and scalable quantum-circuit realization of CTQWs on random graphs. A common algorithm to simulate a general graph (of size $N = 2^n$ for $n$ qubits) on a quantum circuit is based on Pauli decomposition of the graph Hamiltonian, which can yield $O(4^n)$ terms, and require $O(N^2\log N)$ time for coefficient computation. In contrast, our LPA uses $O(2^n)$ terms, in $O(N^2)$ time. Our circuit provides a graph-agnostic framework for CTQWs, implemented via a Trotter–Suzuki product formula and confirming error scaling consistent with theoretical Trotter error bounds. To further test the circuit performance, we study the localization behavior of the CTQW. In our case, localization originates from Laplacian spectral degeneracies rather than disorder (Anderson-type), and our circuit faithfully reproduces these localization phenomena and spectral structure for a random graphs with high accuracy.
\end{abstract}

\maketitle

\section{Introduction}

Quantum computers provide a natural framework for simulating quantum dynamical processes that are otherwise challenging for classical computation~\cite{Feynman1982, Lloyd1996, Georgescu_RevModPhys_2014, Fauseweh2024}. Within this context, quantum walks (QWs) have emerged as powerful and versatile tools~\cite{farhi1998quantum, childs2002example, Childs_ACM_2003, Manouchehri2014, Ray_PRE2022, Ray_PRE2025}, serving as algorithmic building blocks for graph-based problems~\cite{Aharonov_ACM_2001, Childs_ACM_2003, Kempe01072003}, providing a rich framework for modelling quantum transport~\cite{MULKEN201137}, and probing complex networks~\cite{Mukai_PRR_2020}. As quantum analogues of classical random walks, QWs leverage superposition and interference in place of stochasticity resulting in ballistic spreading, localization, and rich dynamical behaviors~\cite{Kempe01072003, Portugal2013, Manouchehri2014, inui2004localization, chakraborty2020fast}. Their versatility has led to applications ranging from quantum optimization and simulation to studies of energy transfer, topological phases, and transport in complex networks~\cite{Mohseni2008, KitagawaPRA2010, MULKEN201137, Mukai_PRR_2020, Duda_PRR_2023, Campos2023, LeePRD2025, HuertaAlderete2020}. 

Quantum walks are broadly classified into discrete-time quantum walk (DTQW)~\cite{Kempe01072003,AharonovPRA1993} and continuous-time quantum walk (CTQW)~\cite{farhi1998quantum}. In DTQWs, evolution proceeds through repeated coin–shift operations, introducing internal degrees of freedom that enable controllability, making them well-suited for circuit design and local graph propagation~\cite{DouglasPRA2009,LokePRA2012,LOKE201764}. In contrast, CTQWs are defined directly on graphs, where the Hamiltonian is typically chosen as the adjacency matrix or the graph Laplacian. They also do not require any extra degree of freedom, such as coin operator.

Implementing CTQWs on quantum circuits remains a fundamental challenge due to the non-local and time-dependent nature of their unitary evolution. However, efficient quantum circuit constructions are essential for CTQWs, since they can bridge theoretical models with practical quantum algorithms and facilitate the exploration of transport phenomena in statistically rich graph topologies, marking a key step toward unifying quantum simulation, computation, and complex network theory~\cite{Aharonov_ACM_2001, Childs_ACM_2003, Kempe01072003, MULKEN201137, Mukai_PRR_2020}.  There are a few existing works that have presented quantum circuits for CTQWs, but they are either for specific classes of degree-regular graphs~\cite{loke2017efficient}, or degree-irregular d-sparse graphs~\cite{ChenPRA2024}. These algorithms implement adjacency-based walks, which are not suitable for degree-irregular Laplacian-based cases. Therefore, for general graphs and random networks, designing a scalable CTQW circuit has remained an open problem, as their continuous evolution depends on the global structure of the graph rather than local connections. 

In this paper, we develop a scalable quantum circuit framework for simulating CTQWs on random graph structures, such as Erd\H{o}s--R\'enyi (ER) random graphs. ER random graphs provide a rich testbed for studying quantum transport and algorithmic robustness since their connectivity can be tuned from sparse to dense regimes. To understand the circuit model, we introduce the \emph{Laplacian partitioning algorithm} (LPA), one of the main outcomes of our paper. We express the CTQW Hamiltonian in terms of the graph Laplacian $\bm{L}$ and apply LPA. We demonstrate that $\bm{L}$ of an $n$-qubit graph can be decomposed into a collection of sparse Laplacians $\{\bm{L}^{(j)}\}$, such that $\bm{L} = \sum_{j=1}^{2^{n}-1} \bm{L}^{(j)}$, where each $\bm{L}^{(j)}$ corresponds to a sparse submatrix of $\bm{L}$. To appreciate the importance of LPA, we recall that quantum circuit implementation for a general graph is a problem of Hamiltonian simulation---which is known to be BQP-complete~\cite{Feynman:85, wocjan2006several, childs2017lecture}, \emph{i.e.,} efficient classical solutions are unlikely. A widely used strategy in this context is the implementation of Trotter–Suzuki~\cite{trotter1959product, suzuki1976generalized, suzuki1991general, childs2021theory} decomposition (TSD) or product formulas~\cite{ChildsPhysRevLett2019}. TSD enables the implementation of the full time evolution operator  $U(t) = \exp({-iHt})$, while partitioning the Hamiltonian ($H$) into a sum of local Hamiltonians (not necessarily commuting with each other). In general, the $2^n \times 2^n$-dimensional $H$, acting on $n$ qubits, is written in terms of a sum of $n$-length Pauli strings. The number of terms for a general Hamiltonian in a Pauli string decomposition can grow up to $O(4^n)$, and its time complexity goes as $O(N^2 \log N)$, for $N=2^n$~\cite{Hantzko_2024, Georges_2025}. Moreover, Pauli decomposition requires \emph{additional time} to group commuting Pauli strings together, adding further computational overhead. In contrast, we apply TSD by decomposing a general graph Hamiltonian following LPA. LPA generates fewer terms $O(2^n)$, in less time $O(N^2)$, saving both memory and time complexity.  

Another key feature of our LPA construction is that each $\bm{L}^{(j)}$ is permutation-similar to a block-diagonal Hamiltonian consisting of $2\times 2$ nontrivial blocks. These permutation matrices have a direct representation in terms of \cn ~gates. We present a scalable quantum circuit construction of the block-diagonal Hamiltonian. The full-time evolution operator $\exp({-iHt})$ is then implemented using a TSD scheme applied to LPA partitioned sub-Laplacians. By combining graph partitioning with quantum circuit synthesis, our method establishes a protocol for simulating CTQWs on arbitrary random graphs. It offers a practical alternative to conventional Hamiltonian simulation techniques as well, thus ensuring a wide applicability of our quantum circuit.

The rest of the paper is organized as follows. In Sec.~\ref{sec:Preliminary}, we review the preliminary concepts of graphs, continuous-time quantum walks (CTQWs), and define localization. Section~\ref{sec:G_part_alg} introduces the graph Laplacian partitioning algorithm, which forms the foundation for constructing quantum circuits for CTQWs. Section~\ref{sec:quant_ckt} is devoted to the design of quantum circuits for CTQWs, while Sec.~\ref{sec:Fid_test} presents their application to CTQW implementations. In Sec.~\ref{sec:loc_res}, we analyze the accuracy of the Trotterized circuit evolution and study localization for CTQW circuit simulations. Finally, Sec.~\ref{sec:conclusion} summarizes our findings and outlines future perspectives.

\section{Theoretical preliminaries}
\label{sec:Preliminary}
\subsection{\label{sec:GS_ctqw}Graphs and Continuous-Time Quantum Walks}

Let $G = (V, E)$ be an undirected graph where $V = \{v_1, v_2, \dots, v_N\}$ denotes the set of vertices and $E \subseteq \{ \{v_i, v_j\} \mid i < j \}$ is the set of undirected edges. The structure of the graph is described by the $N \times N$ adjacency matrix $\bm{A}$~\cite{Ray_PRE2025}, defined as,
\begin{equation}\label{eq:adj_def}
    [\bm{A}]_{ij} =\begin{cases}
            1, & \text{if } \{v_i, v_j\} \in E, \\
            0, & \text{otherwise}.
            \end{cases}
\end{equation}
For undirected graphs, $\bm{A}$ is symmetric. The degree of a vertex $v_i$ is given by $d_i = \sum_{j=1}^{N} A_{ij}$, and the diagonal degree matrix $\bm{D}$ is defined by $[\bm{D}]_{ii} = d_i$. The Laplacian matrix of the graph is then $\bm{L} = \bm{D}-\bm{A}$~\cite{Ray_PRE2025}.

In contrast to undirected simple graphs with predefined edges, random graphs~\cite{Bollobás_2001} are generated by probabilistic rules and can be viewed as a collection of vertices with edges chosen at random. A widely studied class of random graphs is the Erd\H{o}s-R\'enyi random graph (ERG) $G(N, p)$~\cite{erdos1959random, erdos1959evolution}, where each possible edge between $N$ vertices is included independently with probability $p$. In this study, we use ERG for constructing our quantum circuit algorithms for continuous-time quantum walks~\cite{farhi1998quantum}, as ERG offers a generic and statistically well-defined model and is widely studied in the literature as well~\cite{chakraborty2020fast,tindall_2022_Quantumphysics,dutta_2025_Discretetimeopen}. The adjacency matrices of the ERG are symmetric, with an expected vertex degree $\langle d \rangle = p(N-1)$. So, for low values of $p$, the ERG becomes sparse. The structural randomness of these graphs ensures that the successful performance of our quantum circuit algorithm on them generalizes to a broad class of graphs, thereby providing a robust and meaningful testbed for our methods.

A given graph can be mapped onto a quantum system by defining a Hamiltonian that reflects the connectivity of the graph~\cite{chung1997spectral, childs2002example,loke2017efficient,Ray_PRE2022}. Two common choices of graph Hamiltonians are,

\begin{align}
H &= -\gamma \bm{A} \quad \text{(adjacency-based)},\label{eq:hamil_ad} \\
\text{or,}& \notag \\
H &= - \gamma \bm{L} = - \gamma(\bm{A} - \bm{D}) \quad \text{(Laplacian-based)}, \label{eq:hamil_lap}
\end{align}

where $\gamma$ is the uniform hopping rate, denoting the transition probability per unit time between any two connected vertices. For regular graphs~\cite{chung1997spectral, Bollobás_2001} where each vertex has the same degree, the Hamiltonians in Eqs.~\eqref{eq:hamil_ad} and~\eqref{eq:hamil_lap} generate equivalent dynamics up to a global phase~\cite{loke2017efficient}. However, for irregular graphs, this equivalence no longer holds. In such cases, the degree matrix $\bm{D}$ is not proportional to the identity, so the eigenvalue shifts it introduces cannot be factored out as a constant phase in the evolution operator $\exp{(-i\bm{A}t)}$. Instead, they modify the relative phases of the eigenstates, producing distinct oscillatory behavior. To maintain interpretational consistency, and since we are dealing with random graphs, we will adhere to Laplacian-based Hamiltonians Eq.~\eqref{eq:hamil_lap}. This choice allows us to capture the structural inhomogeneity of the underlying graph more accurately during CTQW evolution.

The dynamics of quantum systems evolving over graph structures are elegantly captured by the framework of continuous-time quantum walks (CTQWs)~\cite{farhi1998quantum,Ray_PRE2022,Ray_PRE2025}. The system evolves in a Hilbert space ${\cal H}$ of $N$-dimension, with $\ket{\psi(t)}$ representing the state at a given time $t$. ${\cal H}$ is spanned by the computational basis $\set{\ket{j}}_{j=0}^{N-1}$, with each basis state $\ket{j}$ corresponds to vertex $v_j$. The state of the system at time $t$ is represented by a quantum state $\ket{\psi(t)}$ in a Hilbert space of $N$ dimensions, where the basis states $\ket{j} = \{\ket{0}, \ket{1}, \dots, \ket{N-1}\}$ correspond to the vertices $\{v_0, v_1, \dots, v_{N-1}\}$ of the graph. The probability of finding the walker at vertex $v_j$ at time $t$ is given by,

\begin{equation}\label{eq:prob_walk}
    p_{j}(t) = |{\bra{j}\ket{\psi(t)} }|^2.
\end{equation}

The time evolution of the state is governed by the Schr\"{o}dinger equation ($\hbar =1$),
\begin{equation}
i \frac{d}{dt} \ket{\psi(t)} = H \ket{\psi(t)},
\label{eq:schrodinger_ctqw}
\end{equation}
with formal solution,
\begin{equation}\label{eq:evolution}
    \ket{\psi(t)} = e^{-i H t} \ket{\psi(0)}.
\end{equation}

\subsection{Localization}
\label{sec:localization}

In quantum walks, the localization implies non-vanishing probability of finding the walker at, or near its initial position in the long-time limit~\cite{inui2004localization, chakraborty2020fast, mandal2022limit}. This localization can arise not only from disorder (Anderson-type localization~\cite{anderson1958, lahini2008anderson, crespi2013anderson, vakulchyk2017anderson}) but also from structural features of the graph, such as symmetries or spectral degeneracies~\cite{inui2004localization,bueno2020null,balachandran2024disorder}.

Let the walker initially occupy a vertex $v_0$, with state $\ket{\psi}$. To characterize the long-time behavior of the walker at a given vertex, we define the \emph{time-averaged probability} at vertex $v_j$ as (using Eq.~\eqref{eq:prob_walk}),

\begin{equation}\label{eq:loc_def}
\begin{split}
\overline{p}_C(j) &= \lim_{T \to \infty} \frac{1}{T} \int_0^T p_j(t), \\
 &= \lim_{T \to \infty} \frac{1}{T} \int_0^T |\bra{j} e^{-i H t} \ket{\psi} |^2 \, dt.
\end{split}
\end{equation}
For a walk with uniform probability distribution (\emph{i.e.,} maximally mixed walk) on a graph with $N$ vertices, the walker is found at each vertex with probability $1/N$. We say that a CTQW exhibits \emph{localization} at a given vertex $v_j$ if, in the long-time limit, the probability of finding the walker at $v_j$ remains strictly greater than $1/N$. In other words,

\begin{equation}
    \overline{p}_C(j)>N^{{-}1}.
\end{equation}

\section{Partitioning the Laplacian}
\label{sec:G_part_alg}

\begin{figure}[ht]
    \centering
    \includegraphics[width=\columnwidth]{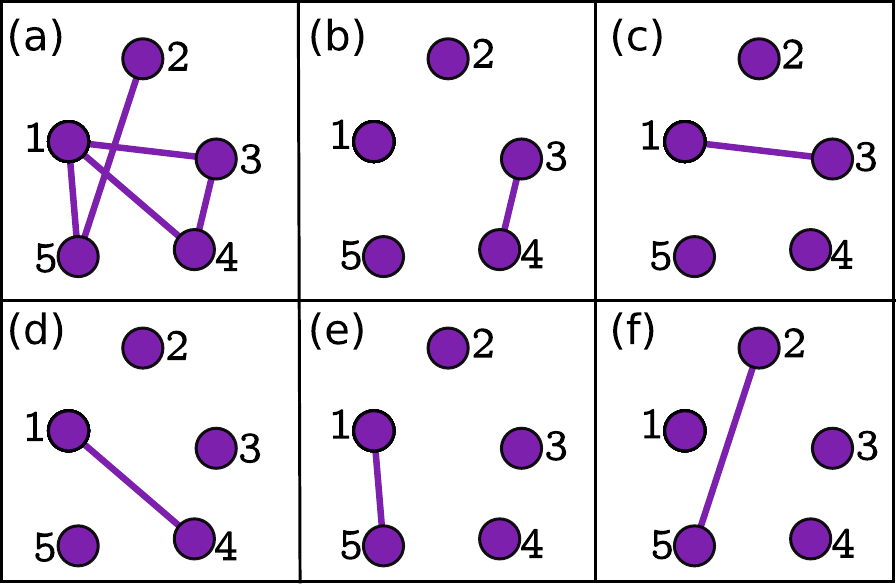}
     \caption{
        (a) A graph $G(N, p)$ with $N = 5$. (b--f) Decomposition of the original graph into subgraphs, each corresponding to a distinct $2$-sparse Hamiltonian representation.
    } \label{fig:gr_partition}
\end{figure}

We now establish the foundation for constructing scalable quantum circuits to simulate CTQWs by introducing a graph Laplacian partitioning algorithm (Fig.~\ref{fig:gr_partition}). In essence, the partitioning algorithm allows us to break the Laplacian of a given graph into a set of Laplacians which represent sparse graphs. This is an essential step in our quantum circuit design. The  LPA proceeds in two key stages (i) the generation and indexing of permutation matrices, and (ii) the subsequent breakdown of the Laplacian into a sum of sparse sub-matrices which are permutation-similar to a block-diagonal matrix with $2\times 2$ non-trivial blocks. The primary idea behind the algorithm is to keep track of the position of non-zero elements in the Laplacian operator. We begin with the following definitions.

\begin{definition}\label{def:suppoert}
    The support set $\Gamma_M$ of a $d\times d$ matrix $\bm{M}$ is defined as the set of positions of its non-zero elements. 
\end{definition} 
For example, suppose that a matrix $\bm{A}$ of size $d\times d$ has nonzero elements at positions $i,j = (1,1)$, $(1,d)$, and $(d,1)$. Then, the support set is given by $\Gamma_A = \set{(1,1),(1,d),(d,1)}$.

\begin{definition}\label{def:supportM}
Let $\bm{A}$ be any matrix. Its \emph{support matrix}, denoted $\tilde{\bm{A}}$, is the binary matrix entry-wise defined by
\begin{equation}
[\tilde{\bm{A}}]_{ij}=
\begin{cases}
1,& A_{ij}\neq 0,\\
0,& A_{ij}=0.
\end{cases}
\end{equation}
\end{definition}
In other words, $\tilde{\bm{A}}$ encodes the pattern of zero and nonzero entries \emph{i.e.,} structure of $\bm{A}$.

\begin{definition}\label{def:structsim}
We call two $d\times d$ matrices $\bm{E}$ and $\bm{F}$ structurally similar, $\bm{E}\struct \bm{F}$, if their support sets  $\Gamma_E$ and $\Gamma_F$, respectively, are equal. 
\end{definition} 

Consider, $\Gamma_E=\set{(p,q)|[\bm{E}]_{p,q}\neq 0,\text{ and } 1\leq p,q\leq d}$ and $\Gamma_F=\set{(p,q)|[\bm{F}]_{p,q}\neq 0, \text{ and }1\leq p,q\leq d}$, then $\Gamma_E=\Gamma_F$ implies $\bm{E}$ is structurally similar to $\bm{F}$, \emph{i.e.,} $\bm{E}\struct \bm{F}$. This trivially implies, $\bm{\tilde{E}} = \bm{\tilde{F}}$.

With these definitions, we describe how the Laplacian of a graph can be decomposed into sparse components. Let $\mathcal{G} = (V, E)$ be an undirected graph with $N=2^n$ vertices ($n$ denotes number of qubits). Using Eq.~\eqref{eq:hamil_lap}, ${\cal G}$ can be represented by its $N\times N$ Laplacian matrix $\bm{L}$. $\bm{L}$ is symmetric by construction. Our objective is to decompose $\bm{L}$ into a sum of structured submatrices $\bm{L}^{(j)}$ as,

\begin{equation}\label{eq:L_decomp}
    \bm{L} = \sum_{j = 0}^{N - 1} \bm{L}^{(j)},
\end{equation}

Each $\bm{L}^{(j)}$ is permutation-similar to a block-diagonal matrix composed of $2 \times 2$ non-trivial blocks \emph{i.e.,} 
\begin{equation}\label{eq:l_blockD}    
    \bm{L}^{(j)}=\bm{P}_n^j \bm{L}_{\mathsf{BD}}^{(j)}(\bm{P}_n^j)^T.
\end{equation}

Here $\bm{L}_{\mathsf{BD}}^{(j)}$ is the block diagonal matrix comprised of $2\times 2$ non-zero blocks. The permutation matrix $\bm{P}_n^j$ of size $2^n\times 2^n$ acts on the $j^\th$ sub-matrix, and $T$ denotes the transpose operation. We know that any $2\times 2$ complex matrix $\bm{A}\in M_{2}(\mathbb{C})$ can be represented using generators of SU$(2)$, \emph{i.e.,} Pauli basis using the set of Pauli matrices $\mathcal{S}_{P} = \set{I,X,Y,Z}$. The set of $n$-length Pauli strings ($n$ times tensor product of 2$\times$2 matrics composed of Pauli matrices and the identity matrix) \emph{i.e.,} $\mathcal{S}^{(n)}_P=\{\bigotimes_{i=1}^n\sigma_i|\sigma_i\in \mathcal{S}_P,\  1\leq i\leq n\}$ forms a basis for $M_{2^n}(\mathbb{C})$, the set of $2^n\times 2^n$ complex matrices. The set of Pauli strings defined on such a matrix $\bm{A}\in M_{2^n}(\mathbb{C})$ is denoted by $\mathcal{S}_{P}(A)$. 

\begin{lemma}
\label{lem:structsimlem}
Let $\bm{A}=\bigotimes_{i=1}^n A_i \in \mathcal{S}_{I,X}^{(n)} \subset \mathcal{S}_P^{(n)} $ be an $n$-length Pauli string comprising of $I$, and $X$.
Further, $\bm{B}=\bigotimes_{i=1}^n B_i \in \mathcal{S}_P^{(n)}$ be another $n$-length Pauli string such that 
$\begin{cases}
    B_j\in \{X,Y\} \mbox{ if } A_j=X\\
    B_j\in \{I,Z\} \mbox{ if } A_j=I_2.
\end{cases}$ 

Then $\bm{A}\struct \bm{B}$.  
\end{lemma}

\begin{proof} Since, $X \struct Y$, and $I \struct Z$, their Kronecker products are also structurally similar by definition. Hence, $\bm{A} \struct \bm{B}$.
\end{proof}

Following Lemma~\ref{lem:structsimlem}, we can write the support matrix of any $2^n \times 2^n$ Hamiltonian matrix using the binary Pauli basis $\{I, X\}^{\otimes n}$, where $n$ is the qubit number. We define a \emph{support basis} in the following way. For a given Pauli string as described above, we can replace $I \mapsto 0$, and $X \mapsto 1$, such that we can map a string of the form $\set{IXXI} \mapsto \set{0110}$. This can be further identified with a basis in the $2^n$ dimensional computational space ($n$ is both the string length and the number of qubits, to be identified from the context). To further clarify, consider a three qubit ($n=3$) Pauli string $(IXI)$, which we can write as $(010) \equiv \ket{010}$. Now, as discussed above, we identify this with one of the basis elements in the $2^3=8$ dimensional computational space, such as $\ket{010}\equiv\ket{2}$. 

Therefore, to index all such Pauli strings that span the given $n$ qubit description of the support of $\bm{A}$ \emph{i.e.,} $\bm{\tilde{A}}$, we can use the index $j = 0,\dots,2^n-1$.

Consider the three-qubit case as before. All the possible Pauli bases chosen from the set 
\begin{align}\label{eq:j_index}
\begin{split}    
    \mathcal{S}_{I,X}^{(3)} & =\bigcurly{III, IIX, IXI,  IXX, XII, XIX, \\
    & \qquad XXI, XXX},\\
    &\mapsto \bigcurly{000, 001, 010, 011, 100, 101, 110, 111} \\
    &\mapsto \curly{0,1,2,3,4,5,6,7} 
\end{split}
\end{align} 

where $j$ encodes the indices $ 0, 1 \cdots 7$.

Having established the notion of the support basis and its indexing through $j$, we now turn to the construction of the corresponding permutation matrix $\bm{P}_n^j$. Our objective is to represent the permutation operator using \cn~gates. We use \cnsub{p}{q} to identify the positions of the control ($p$) and target ($q$) qubits of the \cn~gates (excluding $j=0,1$). For the target qubit of \cn, we need to identify the qubit associated with the given value of the index $j$ as discussed above (see Eq.~\eqref{eq:j_index}). To ease our computation load, we fix the last ($n^\th$) qubit as control (or target, depending on $j$, see below). We convert $j$ to its binary equivalent \emph{i.e.,} $j_\text{bin}$. Since we are fixing the $n^\th$ qubit, we remove the right-most value from the binary string of $j_\text{bin}$ and call the rest of the string $b$. In the binary string $b$, we record the positions of ones as $\kappa_j$ (from right to left), and form the index set $\kappa = \set{\kappa_j}$\footnote{As an example, consider $n=4$ and $j=5$. Binary equivalent of $j$ \emph{i.e.}, $j_\text{bin} = 0101$. Dropping the right most value from $j_\text{bin}$ gives--- string $b=010$. Thus, $\kappa = \set{1}$.}. Each $\kappa_j \in \kappa$ indicates a target qubit for a \cn~gate with the fixed control qubit being qubit $n^\th$ \emph{i.e.,} \cnsub{n}{n-\kappa_j -1}. However, this sequence of operations is valid for odd values of $j$. For even $j$, we need two extra \cn~gates where the control is $n-\max\kappa-1$ \emph{i.e.,} \cnsub{n-\max\kappa -1}{n}.

The expression for $\bm{P}_n^j$ can now be written as
\begin{widetext}
    \begin{align}\label{eq:Permut_def}
        \begin{split}
        \bm{P}^{{j}}_{n} &= I^{\bigotimes n}, \quad \text{if } j = 0 \text{ or } 1; \\
            &=\begin{cases}
            \prod_{j}  \cnsubM{n}{n-\kappa_j -1}, \quad \text{if $j$ is odd;}\\
            \\
             \prod_{j} \cnsubM{n-\max\kappa -1}{n}~ \cnsubM{n}{n-\kappa_j -1} ~ \cnsubM{n-\max\kappa -1}{n}, \quad \text{if $j$ is even}.
            \end{cases} 
        \end{split}
    \end{align}
\end{widetext}

This product involving \cn~gates is equivalent to a permutation operation; we provide the proof in Appendix~\ref{appA:permutation}.

For example, for $n=4$ and odd $j$, we have,

\begin{align*}
P^{j=5}_{n=4}\footnotemark[1] &:= &~\vcenter{\hbox{\Qcircuit @C=1em @R=1em {
\lstick{1} & \qw       & \qw \\
\lstick{2} & \targ     & \qw \\
\lstick{3} & \qw       & \qw \\
\lstick{4} & \ctrl{-2}  & \qw \\
} }}\\
\vspace{2pt}\\
P^{j=15}_{n=4} &:=& ~ \vcenter{\hbox{\Qcircuit @C=1em @R=1em {
\lstick{1} & \qw        & \qw       & \targ     & \qw\\
\lstick{2} & \qw        & \targ       & \qw       & \qw\\
\lstick{3} & \targ      & \qw       & \qw       & \qw\\
\lstick{4} & \ctrl{-1}  & \ctrl{-2}       & \ctrl{-3} & \qw\\
} }}
\end{align*}

and for even $j$, is even, then we have,
\begin{align*}
P^{j=4}_{n=4} &:=& ~ \vcenter{\hbox{\Qcircuit @C=1em @R=1em {
\lstick{1} & \qw        & \qw           & \qw           & \qw\\
\lstick{2} & \ctrl{2}   & \targ         & \ctrl{2}      & \qw\\
\lstick{3} & \qw        & \qw           & \qw           & \qw\\
\lstick{4} & \targ      & \ctrl{-2}     & \targ         & \qw\\
} }}\\
\vspace{2pt}\\
P^{j=14}_{n=4} &:=& ~ \vcenter{\hbox{\Qcircuit @C=1em @R=1em {
\lstick{1}& \ctrl{3}    & \qw        & \qw          & \targ     & \qw   & \ctrl{3}  & \qw\\
\lstick{2}& \qw         & \qw        & \targ        & \qw       & \qw   & \qw       & \qw\\
\lstick{3}& \qw         & \targ      & \qw          & \qw       & \qw   & \qw       & \qw\\
\lstick{4}& \targ       & \ctrl{-1}  & \ctrl{-2}    & \ctrl{-3} & \qw   & \targ     & \qw\\
} }}
\end{align*}

Each value of $j$ encodes the underlying support basis, which uniquely identifies the structure of the given matrix. Therefore, two matrices $\bm{A}$ and $\bm{B}$ having different support matrices $\bm{\tilde{A}}$ and $\bm{\tilde{B}}$, respectively, will have distinct $j$ values. 

\begin{lemma}\label{lem:notstructsim}
    Let $\supt{A},\supt{B}\in \mathcal{S}_{I,X}^{(n)}\setminus \{I_{2^n}\}$ such that $\supt{A}\neq \supt{B}$ with corresponding support sets $\Gamma_A$ and $\Gamma_B$. Then $\Gamma_A\cap \Gamma_B=\emptyset$ 
\end{lemma}

\begin{proof} We know from Eq.~\eqref{eq:j_index} that, $j \in \set{0,1,\cdots, 2^n -1}$ for a $n$ qubit Pauli string. And from Theorem~\ref{th:SRBBvPauli} for each $j_1, j_2 \in \set{0,1,\cdots 2^n -1 }$ there exist $\bm{P}_n^{j_1}$ and $\bm{P}_n^{j_2}$ such that 
\begin{equation}   
\bm{P}_n^{j_1}\supt{A}\bm{P}_n^{j_1} = \bm{P}_n^{j_2}\supt{B}\bm{P}_n^{j_2} = I^{\otimes(n-1)}\otimes X. 
\end{equation}
Further, from Proposition~\ref{prop:constructeven}, no two permutation matrix $\bm{P}_n^{j_1}$ and $\bm{P}_n^{j_2}$ for $j_1\neq j_2$ share the same $2$-cycles.

Let, $\Gamma_A\cap \Gamma_B\neq \emptyset$. Let's assume there exists at least one common row and column index $p,q$ such that $[\bm{A}]_{p,q}, [\bm{B}]_{p,q} \neq 0$. Since $\bm{P}_n^{j_1}(I^{\otimes(n-1)}\otimes X)\bm{P}_n^{j_1}=\supt{A}$ and $\bm{P}_n^{j_2}(I^{\otimes(n-1)}\otimes X)\bm{P}_n^{j_2}=\supt{B}$, there exists at least one $2$-cycle that is common in both  $\bm{P}_n^{j_1}$ and $\bm{P}_n^{j_2}$ due to our assumption. This leads to a contradiction. Hence, the lemma is proved.
\end{proof}

\begin{corollary}\label{corol:notstructsimcoro}
     Let $\supt{A},\supt{B}\in \mathcal{S}_{I,X}^{(n)}\setminus \{I_{2^n}\}$ such that $\supt{A}\neq \supt{B}$ with corresponding $\Gamma_A$ and $\Gamma_B$. Then for any two Pauli strings $\bm{E}\in \mathcal{S}_{P}(A)$ and $\bm{F}\in \mathcal{S}_{P}(B)$ $\Gamma_E\cap \Gamma_F=\emptyset$. 
\end{corollary}
\begin{proof}
Readily follows from the definition of structural similarity (Definition~\ref{def:structsim}) and Lemma~\ref{lem:notstructsim}.
\end{proof}
\begin{algorithm}[t!]
\caption{Laplacian partition algorithm for $2^n\times 2^n$ Hermitian matrices}\label{algo:decomposer}
\KwIn{A $2^n\times 2^n$ real symmetric matrix $\bm{L}$}
\KwOut{$\bm{L}^{(j)}$ such that $\bm{L}=\sum_{j=0}^{2^n-1}\bm{L}^{(j)}$, where $\bm{L}^{(j)}=\bm{P}_n^j \bm{L}_{\mathsf{BD}}^{(j)} \bm{P}_n^j$ and $\bm{L}_{\mathsf{BD}}^{(j)}$ is $2$-sparse block-diagonal with $2\times 2$ blocks}

\textbf{Provided:}
\begin{enumerate}
    \item A $2^n\times 2^n$ real symmetric matrix $\bm{M}$
    \item $\bm{\tilde{H}} = \mqty(1 & 1 \\ 1 & -1)$
    \item Permutation matrices from the set $\bm{P}_n^j$.
\end{enumerate}
\texttt{Extracting block-diagonal and diagonal elements}\\
\For{$j \gets 0$ \KwTo $2^n - 1$}{
    $\bm{L}^{(j)} \gets \mathbf{0}_{2^n \times 2^n}$\;
}

\For{$k \gets 0$ \KwTo $2^n - 1$}{
    $[\bm{L}^{(0)}]_{k,k} \gets [\bm{L}]_{k,k}$\;
}

\For{$k \gets 0$ \KwTo $2^n - 1$  \KwStep $+2$}{
    $[\bm{L}^{(1)}]_{k,k+1} \gets [\bm{L}]_{k,k+1}$\;
    $[\bm{L}^{(1)}]_{k+1,k} \gets [\bm{L}^{(1)}]_{k,k+1}$\;
}
\texttt{Extracting sub-matrices permutation-similar to $2\times 2$ blocks}\\
\For{$j \gets 2$ \KwTo $2^n - 1$}{
    \For{$u \gets 0$ \KwTo $2^{n-1} - 1$}{
        \If{$j$ is odd}{
            Compute $\alpha_{{\kappa}_{\frac{j-1}{2}}}(u)$ and $\beta_{{\kappa}_{\frac{j-1}{2}}}(u)$\;
            \If{$\alpha < \beta$}{

                $[\bm{L}^{(j)}]_{\alpha-1,\alpha} \gets [\bm{L}]_{\alpha-1,\beta}$\;
                $[\bm{L}^{(j)}]_{\beta,\beta-1} \gets [\bm{L}]_{\alpha,\beta-1}$\;
                $[\bm{L}^{(j)}]_{\alpha, \alpha-1} \gets [\bm{L}]_{\beta, \alpha-1}$\;
                $[\bm{L}^{(j)}]_{\beta-1, \beta} \gets [\bm{L}]_{\beta-1, \alpha}$\;
            }
        }
        \Else{
            Compute $\alpha_{{\kappa}_{\frac{j}{2}}}(u)$ and $\beta_{{\kappa}_{\frac{j}{2}}}(u)$\;
            \If{$\alpha < \beta$}{

                $[\bm{L}^{(j)}]_{\alpha-1,\alpha} \gets [\bm{L}]_{\alpha-1,\beta}$\;
                $[\bm{L}^{(j)}]_{\beta,\beta+1} \gets [\bm{L}]_{\alpha,\beta+1}$\;
                $[\bm{L}^{(j)}]_{\alpha, \alpha-1} \gets [\bm{L}]_{\beta, \alpha-1}$\;
                $[\bm{L}^{(j)}]_{\beta+1, \beta} \gets [\bm{L}]_{\beta+1, \alpha}$\;
            }
        }
    }
}
\end{algorithm}

It can be observed that, the elements of $\mathcal{S}_{\{I,X\}_{j}}^{(n)}$ for $j\in \{1,\hdots,2^{n}-1\}$ is permutation similar, \emph{i.e.}, $\mathcal{S}_{\{I,X\}_{j}}^{(n)} = \bm{P}_n^j (I_2^{\otimes (n-1)}\otimes X) \bm{P}_n^j$~(see Appendix~\ref{appA:permutation_sim_Pauli}). Please note that the total number of Pauli strings in a $n$-qubit system is $4^n$, the number of similar structural sets is $2^n -1$.

Thus, we get
\begin{equation}
    \bm{L}=\sum_{j=0}^{2^n-1}\bm{P}_n^j\bm{L}_{\mathsf{BD}}^{(j)}\bm{P}_n^j
\end{equation} 

Since $\bm{P}_n^j$ is symmetric and orthogonal--- $\bm{P}_n^j=(\bm{P}_n^j)^T$. Thus, from our discussions so far, it is evident that $\bm{L}_{\mathsf{BD}}^{(j)} = \bm{P}_n^j \bm{L}^{(j)} \bm{P}_{n}^j$ is a $2$-sparse (each row and column have at most $2$ non-zero elements) block-diagonal matrix with $2\times 2$ non-trivial blocks. Despite its apparent simplicity, generating all $\bm{L}_{\mathsf{BD}}^{(j)}$ involves a sequence of consecutive matrix multiplications, which can become computationally demanding.

In order to lower the complexity of the algorithm, we exploit the sparse structure of $\bm{P}_n^j$. For some real symmetric matrix $\bm{M}$, we denote $\bm{M}^\prime =\bm{P}_n^j\bm{M}\bm{P}_n^j$. When $j$ is odd, one can observe from Proposition~\ref{prop:constructeven} in Appendix~\ref{appA:permutation}, that 
\begin{align}
    \begin{split} 
       [\bm{M}^\prime ]_{\scalebox{0.7}{${\alpha_{\kappa_{\frac{j-1}{2}}}(u)-1, \alpha_{\kappa_{\frac{j-1}{2}}}(u)}$}} &= [\bm{M}]_{\scalebox{0.7}{${\alpha_{\kappa_{\frac{j-1}{2}}}(u)-1, \beta_{\kappa_{\frac{j-1}{2}}}(u)}$}} \\
       [\bm{M}^\prime ]_{\scalebox{0.7}{${\beta_{\kappa_{\frac{j-1}{2}}}(u), \beta_{\kappa_{\frac{j-1}{2}}}(u)-1}$}} &= [\bm{M}]_{\scalebox{0.7}{${\alpha_{\kappa_{\frac{j-1}{2}}}(u), \beta_{\kappa_{\frac{j-1}{2}}}(u)-1}$}}.
    \end{split}
\end{align} 

Subsequently, if $j$ is even, then
\begin{align}
    \begin{split} 
        [\bm{M}^\prime ]_{\alpha_{\kappa_{\frac{j}{2}}}(u)-1, \alpha_{\kappa_{\frac{j}{2}}}(u)} &= [\bm{M}]_{\alpha_{\kappa_{\frac{j}{2}}}(u)-1, \beta_{\kappa_{\frac{j-1}{2}}}(u)} \\
        [\bm{M}^\prime ]_{\beta_{\kappa_{\frac{j}{2}}}(u), \beta_{\kappa_{\frac{j}{2}}}(u)+1} &= [\bm{M}]_{\alpha_{\kappa_{\frac{j}{2}}}(u), \beta_{\kappa_{\frac{j-1}{2}}}(u)+1},
    \end{split}
\end{align}.

where $u$ is an integer such that $0\leq u\leq 2^{n-1}-1$ and have a corresponding binary representation $u=(u_{n-2},\hdots,u_0)$. The symbols $\kappa,\alpha,\beta$ are as introduced in Proposition~\ref{prop:constructeven} and also in Ref.~\cite{sarkar2024scalable}. $\bm{M}$ being an symmetric matrix one can easily observe that after performing permutation if $[\bm{M}]_{i,j} \to [\bm{M}]_{i', j'}$ then $[\bm{M}]_{j, i} \to [\bm{M}]_{j', i'}$. 

Thus, we can directly substitute matrix multiplication with swapping the elements around by harnessing the sparsity pattern of the permutation matrices. We finally arrive at our decomposition Algorithm~\ref{algo:decomposer}. 

\begin{theorem}
 The running time complexity for Algorithm~\ref{algo:decomposer} is $O(N^2)$ where $N=2^n$.   
\end{theorem}
\begin{proof}
    Follows from the Algorithm immediately.
\end{proof} 

\section{Quantum Circuit decomposition}
\label{sec:quant_ckt}

For a given graph, we simulate the time evolution operator $U = \exp(-i H \delta t)$, where $H = -\gamma \bm{L}$ Eq.~\eqref{eq:hamil_lap}. Using Eqs.~\eqref{eq:L_decomp},~\eqref{eq:l_blockD} the Hamiltonian can be expressed as,
\begin{equation}\label{eq:hamil_BD_decomp}
    H= -\gamma \sum_{j=0}^{2^n-1}\bm{P}_n^j \;\bm{L}_{\mathsf{BD}}^{(j)}\; \bm{P}_n^j.
\end{equation} 
To simulate the corresponding dynamics on a quantum circuit, we approximate the unitary evolution operator $U$ via a first-order Trotter~\cite{trotter1959product, suzuki1976generalized, suzuki1991general, childs2021theory} expansion. The effective unitary becomes,

\begin{align}\label{eq:uni_BD_decomp}
    \begin{split}
        U &= \exp(i \gamma \sum_{j=0}^{2^n-1}\bm{P}_n^j \;\bm{L}_{\mathsf{BD}}^{(j)}\; \bm{P}_n^j \delta t)\, ,  \\
        & = \prod_{j=0}^{2^n -1} \bm{P}_n^j \exp(i \gamma \;\bm{L}_{\mathsf{BD}}^{(j)}\; \delta t)  \bm{P}_n^j\, + \varepsilon_{\delta t}, \\
        &= \prod_{j=0}^{2^n -1} \bm{P}_n^j \Breve{\bm{U}}_{\mathsf{BD}}^{(j)} \bm{P}_n^j\, + \varepsilon_{\delta t}
    \end{split}
\end{align}

where $\varepsilon_{\delta t}$ is first-order Trotter–Suzuki bound (for details see section~\ref{sec:Trotter_error}). We define $\Breve{\bm{U}}_{\mathsf{BD}}^{(j)} = \exp(i \gamma \bm{L}_{\mathsf{BD}}^{(j)} \delta t)$ as the block-diagonal unitary corresponding to the $j^{\text{th}}$ component (for $\Breve{\bm{U}}_{\mathsf{BD}}^{(0)}$ is a diagonal unitary matrix).

Now, for  each block-diagonal unitary $\Breve{\bm{U}}_{\mathsf{BD}}^{(j)}$ where $j\neq 0$, we seek circuit decomposition. Since $\Breve{\bm{U}}_{\mathsf{BD}}^{(j)}$ is composed of $2\times 2$ non-trivial blocks, we can write the following,

\begin{widetext}
\begin{equation}\label{eq:uni_BD_mat}
    \Breve{\bm{U}}_{\mathsf{BD}}^{(j)} = 
    \mqty(
        \Breve{\bm{U}}_{1}^{(j)}(\theta_1, \zeta_1, \varphi_1) &  &  &  \\
        &  \Breve{\bm{U}}_{2}^{(j)}(\theta_2, \zeta_2, \varphi_2)&  &  \\
        &  &  \ddots &  \\
        &  &  & \Breve{\bm{U}}_{2^{n-1}}^{(j)}(\theta_{2^{n-1}}, \zeta_{2^{n-1}}, \varphi_{2^{n-1}}) \\
    ),
\end{equation}
\end{widetext}
where each $2\times 2$ block is of the form

\begin{align}
    \begin{split}
    &\Breve{\bm{U}}_\mathsf{b}^{(j)}(\theta_b, \zeta_b, \varphi_b)
    \\
    &= \left(\begin{smallmatrix}
        \exp{i(\theta_b + \zeta_b)} \cos\varphi_b &  \exp{i(\theta_b - \zeta_b)} \sin\varphi_b \\
        -\exp{i(- \theta_b + \zeta_b)} \sin\varphi_b &  \exp{i(- \theta_b - \zeta_b)} \cos\varphi_b 
    \end{smallmatrix}\right)\in \text{SU}(2)
    \end{split}
\end{align}

and

\begin{equation}\label{eq:U_tild_BD}
    \Breve{\bm{U}}_{\mathsf{BD}}^{(0)}=\exp(i \gamma \bm{L}^{(0)} \delta t) =\bigoplus_{l=1}^{2^{n-1}} \text{diag}(e^{i\Delta_{b'}^{(l)}}, e^{i\Delta_b^{(l)}})
\end{equation}
Here $\Delta_b, \Delta_{b'}, \theta_b, \zeta_b$, and $\varphi_b$ are real-valued parameters. 

To understand the circuit-level realization of $\left( \Breve{\bm{U}}_{BD} \right)$, we fix the $n^{\text{th}}$ qubit as the \emph{target}, with all remaining qubits acting as \emph{controls}. For an axis $a\in \set{Y,Z}$ we use the standard one-qubit rotations 
\begin{equation}
R_y(\theta)= \mqty(\cos\theta & \sin\theta  \\ -\sin\theta & \cos\theta ),
\end{equation}
and 
\begin{equation} 
R_z(\theta) = \mqty(e^{i\theta} & 0  \\ 0 & e^{-i\theta}).
\end{equation}
We write $\cnsubM{c}{t}$ for a \cn~ with control $c$ and target $t$, and it can be expressed as,
\begin{equation}\label{eq:cnot-projector}
    \cnsubM{c}{t} = \ket{0}\bra{0}_{c} \otimes I_t \;+\; \ket{1}\bra{1}_{c} \otimes X_t.
\end{equation}

Two basic conjugation identities that will be used are, 
\begin{align}\label{eq:XRX}
    \begin{split}
        &X\,R_a(\phi)\,X = R_a(-\phi)\:\text{ for }a\in \set{Y,Z}\,,\\
        &R_a(\alpha)\,R_a(\beta)=R_a(\alpha+\beta).
    \end{split}
\end{align}

Since any $2 \times 2$ special unitary matrix has a ZYZ decomposition, $\Breve{\bm{U}}_{BD}$ has a circuit from using the multi-controlled rotation gates, which we explain below.

\begin{definition}
For $n$-qubit systems, let $n \geq 2$. The first $n-1$ qubits form the control register, and the $n$-th qubit is the target. For a list of angles $\Theta=\{\theta_b\}_{1 \leq b \leq 2^{n-1}}$ the $n$-qubit multi-controlled rotation around axis $a$ is the block-diagonal unitary defined as \cite{mikko2004quantumcircuits, mikko2005transformation, sarkar2023scalable}, 

\begin{equation}
    F_n(R_a;\Theta) = \mqty(
    R_a(\theta_1)&   &  \\
    & \ddots  &   \\
    &    &R_a(\theta_{2^{n-1}})
    )
\end{equation}
\end{definition}

The corresponding circuit is given below, 

\begin{eqnarray}\label{circ:ZYZ_decomp}
    {\Qcircuit @C=1em @R=.7em {
        &\lstick{1}&\qw& \gate{} &\qw\\
        &\lstick{\vdots}&\qw&\gate{ }\qwx[-1]&\qw\\
        &\lstick{n-1}&\qw&\gate{ }\qwx[-1]&\qw\\
        &\lstick{n}&\qw&\gate{F_n(R_a)}\qwx[-1]&\qw\\}}
\end{eqnarray}

Example: The 2-qubit multi-controlled rotation gate can be implemented using the circuit in Eq.~\eqref{circ:ZYZ_decomp} as,
\begin{equation}
    \mbox{\scalebox{0.8}{
    \Qcircuit @C=1.4em @R=.9em {
        \lstick{1} &  \qw  & \ctrl{1} &  \qw  & \ctrl{1} & \qw \\
        \lstick{2} & \gate{R_{a}(\theta_{1})} & \targ{} & \gate{R_{a}(\theta_{2})} & \targ{} & \qw }}}
\end{equation}

The $3$-qubit multi-controlled rotation gate circuit from circuit~\ref{circ:ZYZ_decomp} is,
\begin{widetext}
    \begin{equation}\label{circ:ZYZ_decomp3}
    \Qcircuit @C=1.4em @R=.9em {
    \lstick{1} & \qw & \qw       & \qw & \qw       & \ctrl{2} & \qw       & \qw       & \qw & \qw       & \ctrl{2} & \qw \\
    \lstick{2} & \qw & \ctrl{1}  & \qw & \ctrl{1}  & \qw      & \qw       & \ctrl{1}  & \qw & \ctrl{1}  & \qw      & \qw \\
    \lstick{3} & \gate{R_z(\theta_{1})} & \targ & \gate{R_z(\theta_{2})} & \targ & \targ & \gate{R_z(\theta_{3})} & \targ & \gate{R_z(\theta_{4})} & \targ & \targ & \qw
}
    \end{equation}
\end{widetext}

\begin{definition}
    A unitary $\bm{U}$ on $n$-qubits is $2 \times 2$ block diagonal with,
    \begin{equation}\label{eq:2sparse}
        \bm{U}=\bigoplus_{1\leq b \leq 2^{n-1}} \bm{U}_b, \;\; \text{ and }\;\; \bm{U}_b\in \text{SU}(2)
    \end{equation}  
\end{definition}

Every $\bm{U}_b$ admits a ZYZ factorization
    \begin{equation}\label{eq:zyz-per-block}
         \bm{U}_b=R_z(\alpha_b)\,R_y(\gamma_b)\,R_z(\beta_b).    
    \end{equation}
Consequently,
\begin{proposition}\label{prop:zyz}
Any $\bm{U}$ (Eq.~\eqref{eq:2sparse}) can be implemented as \cite{mikko2004quantumcircuits, mikko2005transformation, sarkar2023scalable}

\begin{equation}
    \bm{U} = F_n(R_z;\{\alpha_b\})\;
        F_n(R_y;\{\gamma_b\})\;
        F_n(R_z;\{\beta_b\})\; .
\label{eq:global-zyz}
\end{equation}
\end{proposition}
        
The corresponding circuit representation is given by,

\begin{widetext}
\begin{equation}\label{eq:multi_ctrl_rot}
\Qcircuit @C=1.4em @R=.9em {
\lstick{1}        & \gate{ }\qwx[1] & \qw & \gate{ }\qwx[1] & \qw & \gate{ }\qwx[1] & \qw \\
\lstick{2}        & \gate{ }\qwx[1] & \qw & \gate{ }\qwx[1] & \qw & \gate{ }\qwx[1] & \qw \\
\lstick{\vdots}   & \vdots     &     & \vdots     &     & \vdots     &     \\
\lstick{n-1}      & \gate{ }\qwx[1] & \qw & \gate{ }\qwx[1] & \qw & \gate{ }\qwx[1] & \qw \\
\lstick{n}        & \gate{F_n(R_z(\alpha_1,\ldots,\alpha_{2^{n-1}}))} & \qw
                  & \gate{F_n(R_y(\gamma_1,\ldots,\gamma_{2^{n-1}}))} & \qw
                  & \gate{F_n(R_z(\beta_1,\ldots,\beta_{2^{n-1}}))}  & \qw }
\end{equation}
\end{widetext}

\begin{lemma}\label{lem:angle_parity} \cite{mikko2004quantumcircuits, mikko2005transformation}
For an $n$-qubit circuit, let $k=n-1$ be the number of control qubits and let the last qubit be the target. Consider a sequence of single-qubit rotations $R_a(\omega_1), R_a(\omega_2),\dots,R_a(\omega_{2^{n-1}})$ on the target with $\Omega=\{\omega_i\}_{1 \leq i \leq 2^{n-1}}$, with $a\in \set{Y,Z}$. For each $i$, let $m_i\in\set{0,1}^k$ encode which control lines are connected to the target immediately before $R_a(\omega_i)$.
Then the total unitary is block diagonal in the control basis,
\begin{equation}
    \bm{U}=\bigoplus_{c\in\set{0,1}^k} R_a(\eta_c),
\end{equation} with the block angle for control string $c$ given by

\begin{align}
\begin{split}
    &\eta_c \;=\; \sum_{i=1}^{2^{n-1}} (-1)^{\langle c,m_i\rangle}\,\omega_i,\\
    &\langle c,m_i\rangle \;=\; \bigg(\sum_{j=1}^k c_j m_{i,j}\bigg) (\mathrm{mod}\ 2)\;.
\end{split}
\end{align}
\end{lemma}

The Lemma~\ref{lem:angle_parity} states nothing but a solution of the linear system of equations \cite{mikko2004quantumcircuits, mikko2005transformation, krol2022efficient},
\begin{equation}\label{eq:angle_lin_sys}
    \bm{M}^{\otimes k} \mqty(\omega_1\\
    \omega_2\\
    \vdots\\
    \omega_{2^{k}}) = \mqty(\eta_1\\
    \eta_2\\
    \vdots\\
    \eta_{2^{k}})
\end{equation}

where the matrix elements $[\bm{M}^{\otimes k}]_{ij}$ can be determined using Lemma~\ref{lem:angle_parity} (a detailed discussion is given in Appendix~\ref{appB:Lemma_3}). Eq.~\eqref{eq:angle_lin_sys} is exactly a Walsh-Hadamard transform~\cite{mikko2004quantumcircuits,mikko2005transformation}, where $2^{-k/2} \bm{M}^{\otimes k}$ corresponds to $H^{\otimes k}$. Thus, computing the $\set{\omega_i}$ angles is precisely multiplication by $2^{-k}H^{\otimes k}$ applied to ${\set\eta_i}$ [see Eq.~\eqref{eqB:transformation} for details]. 

As stated earlier (Eq.~\eqref{eq:U_tild_BD}), $\Breve{\bm{U}}_{\mathsf{BD}}^{(0)}$ \emph{i.e.}, $\exp{(iL_0 \delta t)}$, is a diagonal matrix with a special structure $\bigoplus_{j=1}^{2^{n-1}} \text{diag}(e^{i\Delta_{b'}^{(j)}}, e^{i\Delta_b^{(j)}})$. In order to construct quantum circuit for this term, we first decompose $L_0$ into linear combination of diagonal Pauli strings. Note that such strings---composed of only $I$ and $Z$---are easier to compute (partly because we are computing tensor product of vectors instead of matrices), and it can be trivially shown that the concerned decomposition complexity is $O(N^2)$. This decomposition is achieved through Fast Walsh-Hadamard transform~\cite{Georges_2025}. Further, all of the diagonal Pauli strings are commutative and are very easy to implement in a quantum circuit as shown in~\cite{sarkar2023scalable,sarkar2024scalable}. Each diagonal Pauli string requires at most $O(n)$ \cn~gates, and the total number of diagonal strings is $O(N)$. So at most $O(n2^n)$ \cn~gates are required for the unitary diagonal matrix. For every other permutationally similar block-diagonal matrices with $2\times 2$ non-trivial blocks, the total number of \cn~gates required is $O(2^{n-1}+2n)$ ($2^{n-1}$ for block-diagonal and $2n$ for permutations), and there are $2^n-1$ such block-diagonal matrices. Thus, our circuit requires $O(4^{n-1/2}-2^{n-1}+n2^n+n2^{n+1}-2n)=O(4^{n-1/2}+\frac{5}{2}n2^n-2n)\approx O(4^{n})$ \cn~gates; which is an improvement on the number of \cn~gates $O(n4^n)$ required after constructing the circuit following a complete Pauli string decomposition.

Therefore, the final circuit representation of  $\Breve{\bm{U}}_{\mathsf{BD}}$ in Eq.~\eqref{eq:uni_BD_decomp} consists of the ZYZ circuit decomposition (see Eq.~\eqref{eq:multi_ctrl_rot}) and the phase components [\emph{i.e.}, $\Breve{\bm{U}}_{\mathsf{BD}}^{(0)}$  
From the circuit structure of the multi-qubit rotation gate circuit in Eq.~\eqref{circ:ZYZ_decomp} and following the construction of diagonal Pauli-strings in \cite{sarkar2024scalable}, it can be clearly understood that the circuit structure shown above represents the recursive construction of a unitary operator where $n$-qubit circuit structure can be constructed from $n-1$ qubit circuit structure, making the circuit construction scalable.

\section{Continuous-time quantum walk implementation}
\label{sec:Fid_test}

\begin{figure}[t!]
    \centering
    \includegraphics[width=\linewidth]{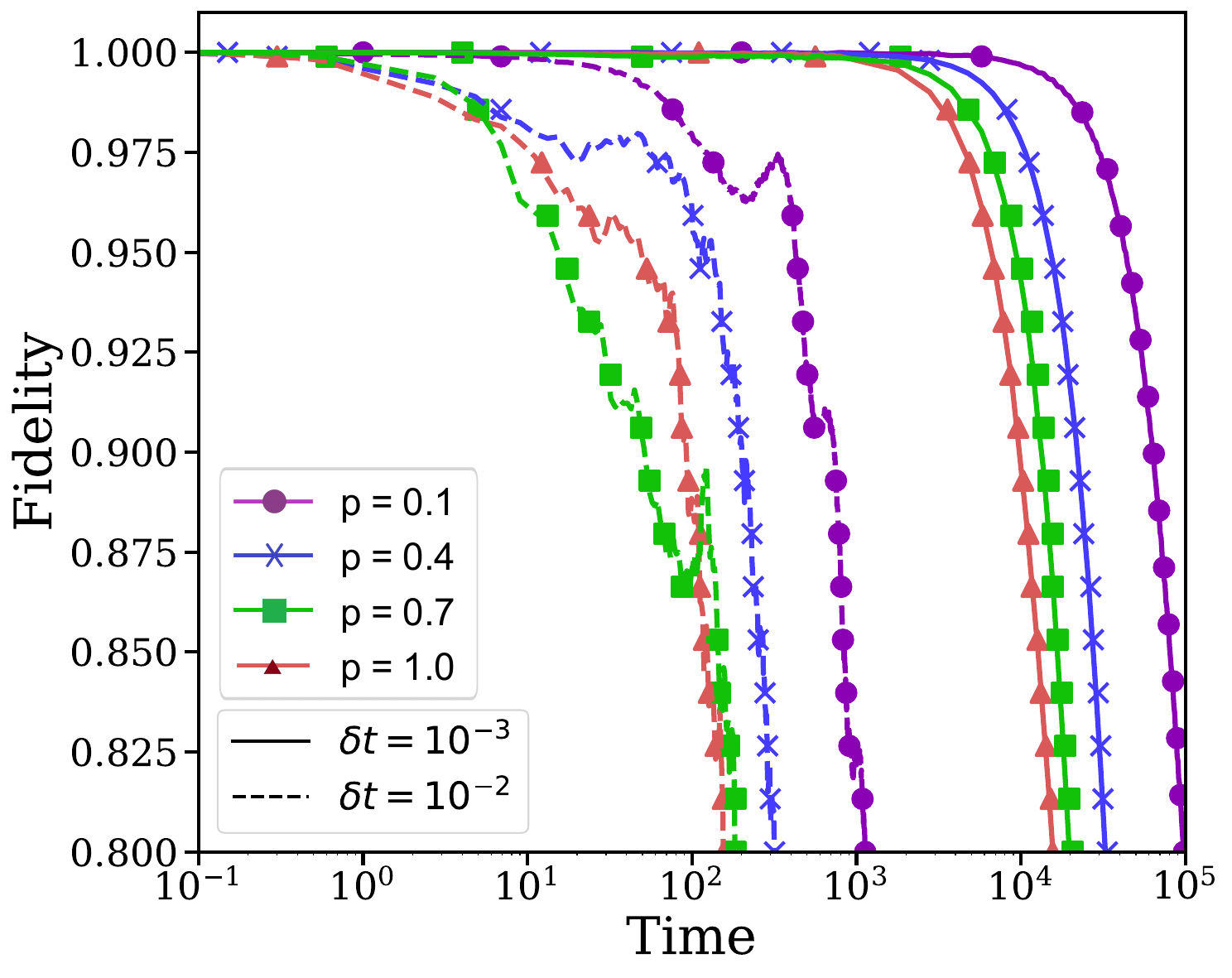}
    \caption{Fidelity plot of the 6-qubit quantum circuit simulating continuous-time quantum walk on Erd\H{o}s-R\'enyi graphs for four different edge probabilities $ p = 0.1, 0.4, 0.7, 1.0 $. The simulation is performed using two Trotter step sizes $ \delta t = 10^{-2} $ (dashed lines) and $ \delta t = 10^{-3} $ (solid lines). Fidelity is computed against the exact unitary evolution operator $ \exp(-iHt) $ using Eq.~\ref{eq:fidelity}. The results demonstrate that smaller Trotter step sizes yield higher circuit fidelity over longer evolution times, with fidelity degrading more rapidly for higher connectivity (larger $p$).}
    \label{fig:cross_fid_vs_time}
\end{figure}

\subsection{Performance of a quantum circuit}

The performance of the quantum circuit, which is outlined in section~\ref{sec:quant_ckt} for the continuous-time quantum walks, is evaluated here. We compare the circuit-evolved states \emph{i.e.} $\ket{\psi_{\mathrm{circuit}}(t)}$ with the state generated by the exact unitary dynamics governed by the Hamiltonian $H = -\gamma \bm{L}$ [Eq.~\eqref{eq:hamil_lap}] \emph{i.e.,} $\ket{\psi_{\mathrm{exact}}(t)}$. The exact evolution is obtained from direct exponentiation of the Laplacian, $\exp(-iHt)$,
 while the circuit dynamics are simulated using a first-order Trotter-Suzuki~\cite{trotter1959product, suzuki1976generalized, suzuki1991general, childs2021theory} decomposition. The fidelity is defined as~\cite{jozsa1994fidelity},
\begin{equation}\label{eq:fidelity}
    F(t) =  \abs{\braket{\psi_{\mathrm{exact}}(t)}{\psi_{\mathrm{circuit}}(t)}}^2,
\end{equation}
This fidelity value quantifies the accuracy of the circuit approximation. Fig.~\ref{fig:cross_fid_vs_time} presents the fidelity against time for a six-qubit system ($N = 2^6$ vertices) performing CTQWs circuit simulation on Erd\H{o}s--R\'enyi graphs with varying edge probabilities $p = 0.1, 0.4, 0.7,$ and $1.0$. The simulations are performed over logarithmically spaced time values up to $t = 10^5$. Two different Trotter step sizes, $\delta t = 10^{-2}$ and $\delta t = 10^{-3}$, are considered to evaluate the circuit performance. 

\begin{figure}[t!]
  \centering
 \subfloat[ \label{fig:fidelity_02} $\delta t = 10^{-2}$]{\includegraphics[width=\columnwidth]{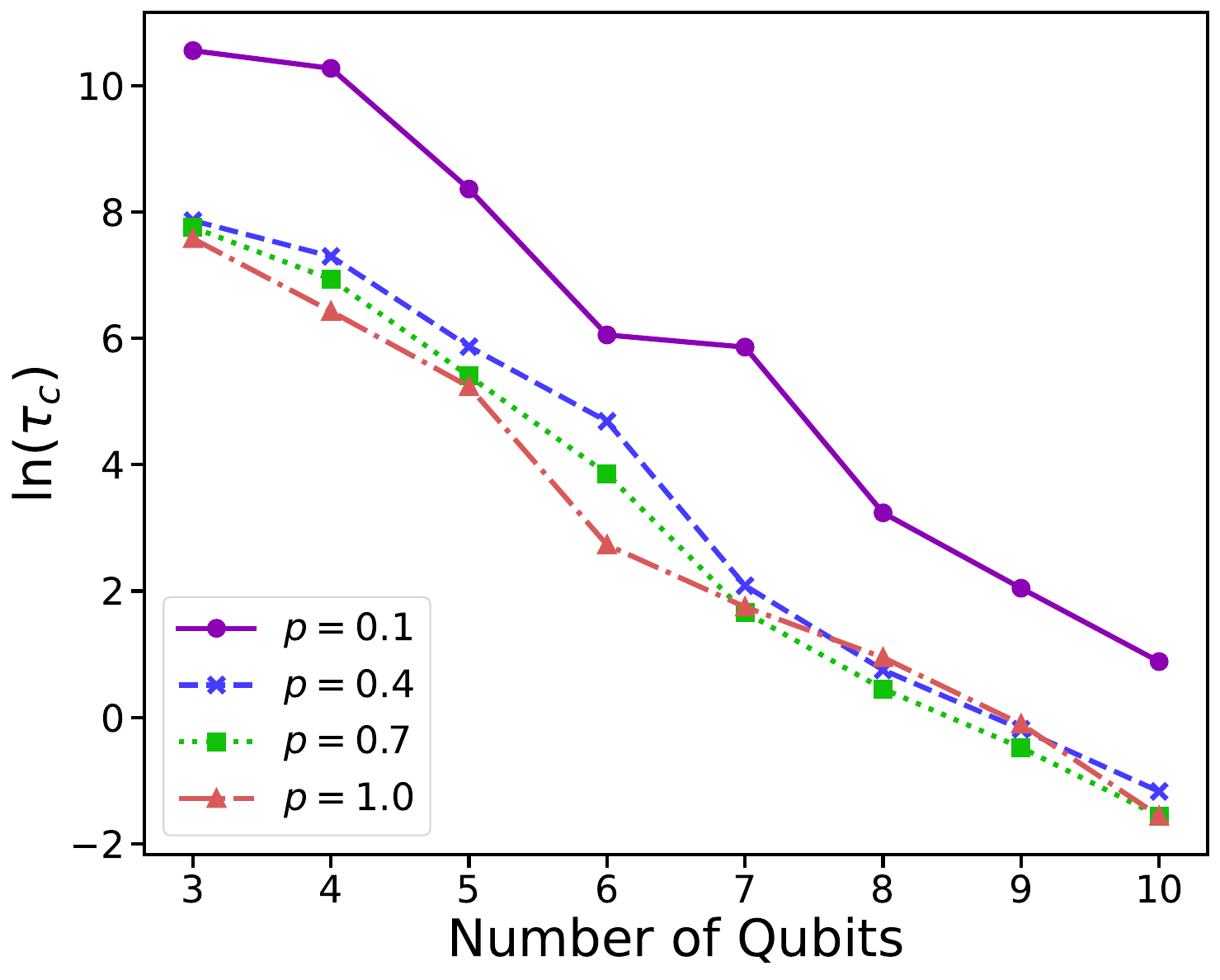}}\\ 
 \subfloat[ \label{fig:fidelity_03} $\delta t = 10^{-3}$]{\includegraphics[width=\columnwidth]{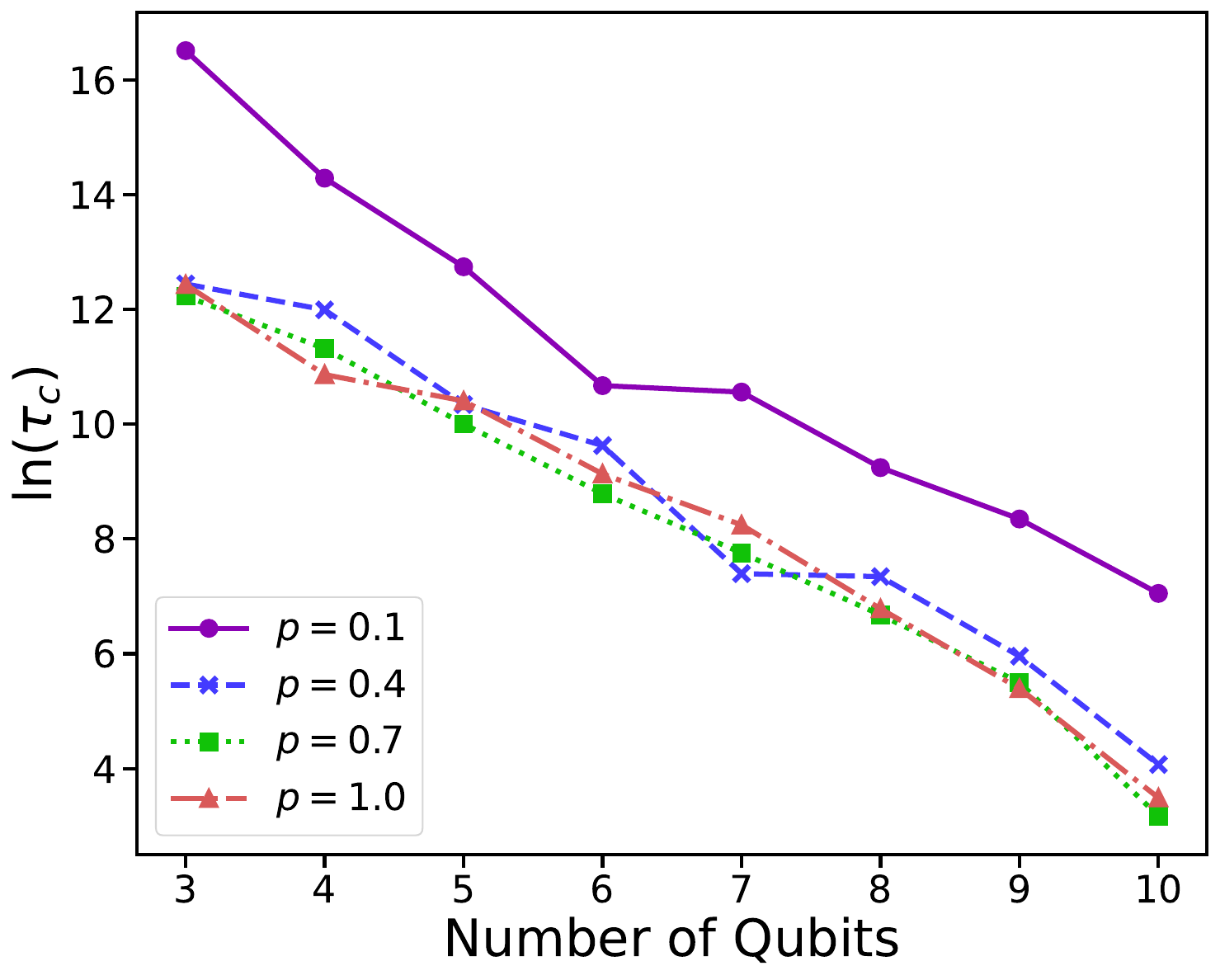}} 
  \caption{Cutoff Time ($\tau_c$) at which the quantum circuit fidelity $\sim 0.95$ plotted against the number of qubits $n$, for different edge probabilities $p$ in the underlying Erd\H{o}s-R\'enyi graph. (a) Results for Trotter time step $\delta t = 10^{-2}$. (b) Same for $\delta t = 10^{-3}$. The fidelity decays more rapidly with increasing number of qubits $n$, and the decay is further for graphs with higher connectivity $p$ and larger Trotter step size $\delta t$.}
  \label{fig:fidelity}
\end{figure}
For all values of $p$, the fidelity degrades over time due to the accumulation of Trotter errors. The results indicate that reducing the Trotter step size improves the accuracy of the simulation. Smaller step sizes ($\delta t = 10^{-3}$) show slow fidelity decay, maintaining $\text{fidelity} > 0.98$ up to $t \sim 10^{4}$. From Fig.~\ref{fig:cross_fid_vs_time}, the dependence of fidelity on graph connectivity is also can be observed---the edge probability $p$ significantly affects the fidelity decay. Sparse graphs exhibit slower fidelity decay because their Hamiltonians contain fewer non-commuting terms. As connectivity $p$ increases, additional non-commutativity accelerates fidelity loss. For fixed $p$, the fidelity remains closer to unity for longer period of time when $\delta t = 10^{-3}$ than when $\delta t = 10^{-2}$. In contrast, for fixed $\delta t$, sparser graphs maintain higher fidelity over longer times. Thus, the departure of fidelity from unity is governed jointly by graph connectivity and the Trotter–Suzuki step size. This behavior is consistent with general results in Hamiltonian simulation, where the Trotter–Suzuki error scales with both the Hamiltonian norm and the chosen time step~\cite{trotter1959product,suzuki1976generalized,suzuki1991general,childs2021theory}.

\subsection{Fidelity scaling}
To further understand the accuracy of our quantum circuit implementation, we analyze the decay of circuit fidelity as a function of system size and graph connectivity. We define the cutoff time $\tau_c$ as the evolution time at which the fidelity drops to approximately $0.95$. The fidelity is averaged over ten independent realizations of Erd\H{o}s–R\'enyi graphs in order to account for statistical fluctuations. The results are shown in Fig.~\ref{fig:fidelity}, where $\tau_c$ is plotted against the number of qubits $n$ for several values of the edge probability $p$. Two Trotter step sizes are considered $\delta t = 10^{-3}$ and $\delta t = 10^{-2}$.

\begin{figure}[t!]
    \centering
    \includegraphics[width=\linewidth]{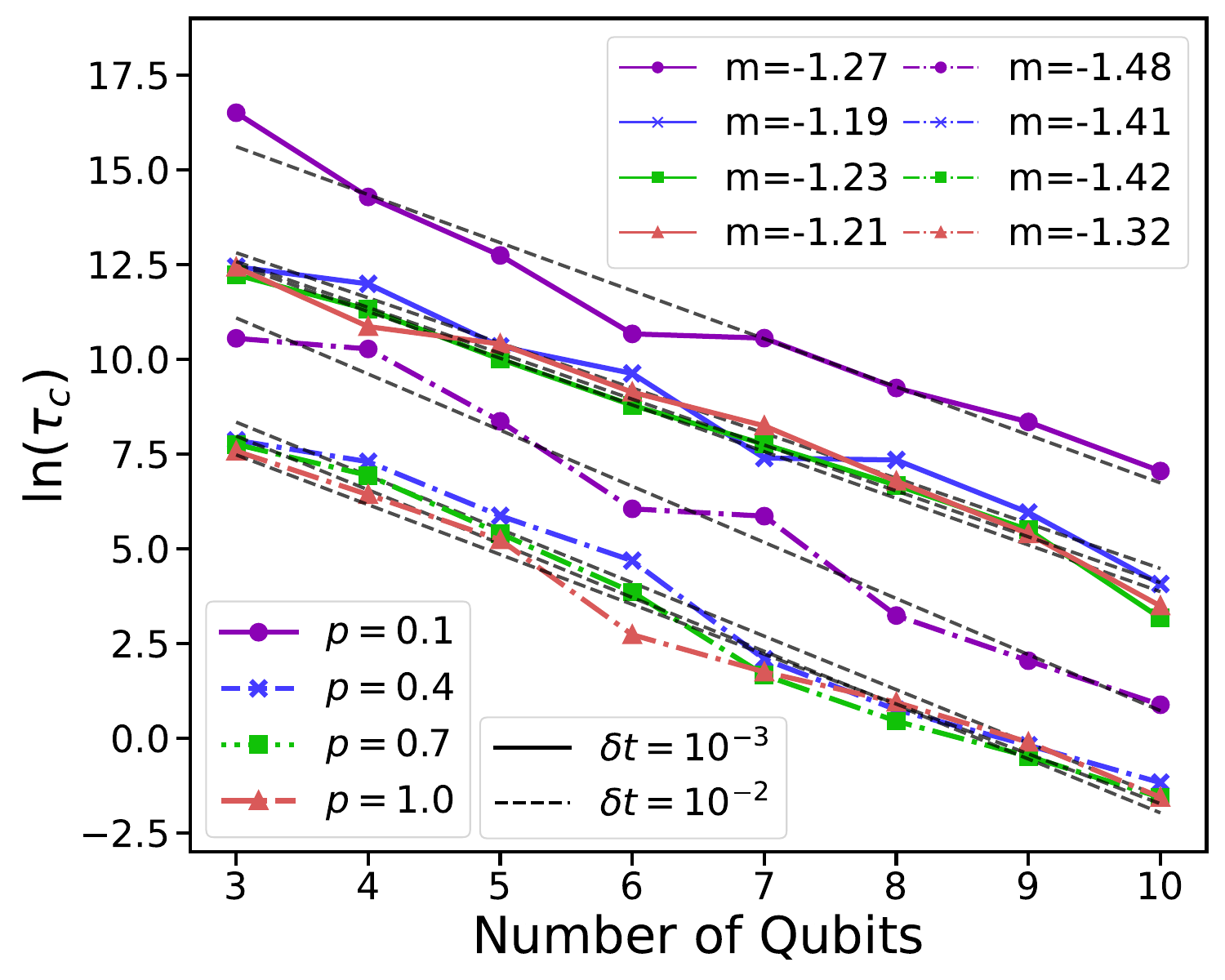}
    \caption{Combined analysis of the cutoff time ($\tau_c$) for fidelity decay (falls below $95\%$) plotted against the number of qubits $n$, including data from both Fig.~\ref{fig:fidelity_02} and Fig.~\ref{fig:fidelity_03}. Each curve corresponds to a different Erd\H{o}s-R\'enyi graph connectivity $p$ and Trotter step size $\delta t$. The straight lines represent exponential fits of the form $T(n) \sim e^{m n + c}$, with fitted slope ($m$) mentioned in the legend.}
    \label{fig:lin_fitcross_time}
\end{figure}
From Fig.~\ref{fig:fidelity} we observe, the cutoff time $\tau_c$ decreases as the number of qubits increases, indicating that the larger the qubit number, the more the number of non-commuting terms, which result in a rapid increase of Trotter errors. Also, the graph connectivity plays an important role, sparse graphs \emph{i.e.}, graphs with low edge probability $p$ depict higher $\tau_c$ than higher $p$ for a fixed number of qubits. Moreover, the choice of Trotter step size significantly affects performance. For $\delta t = 10^{-3}$, $\tau_c$ is larger across all values of $p$ than $\tau_c$ for $\delta t = 10^{-2}$, which indicates that larger Trotter step size $\delta t$ leads to significantly shorter evolution times before fidelity falls below $0.95$.


\subsection{Trotter error}\label{sec:Trotter_error}

\begin{figure}[t!]
    \centering
    \includegraphics[width=\linewidth]{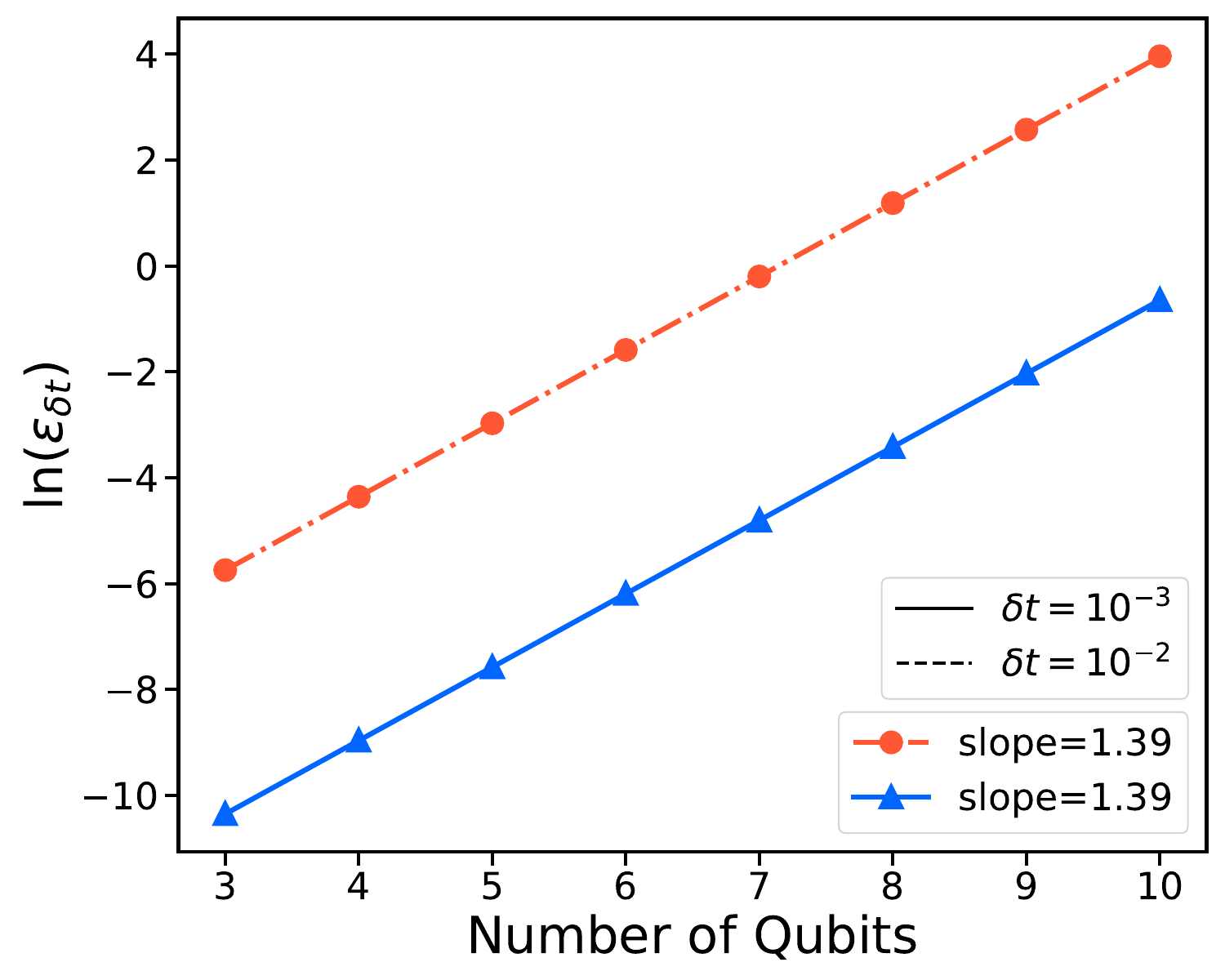}
    \caption{Scaling analysis of the Trotterization error ($\varepsilon_{\delta t}$) at a single Trotter step $\delta t$ as a function of qubit number $n$. Theoretical upper bound of Trotter error ($\varepsilon_{\delta t}$), given by $\delta t^2 \cdot \epsilon \cdot 2^{2n - 1}$, is also fitted with straight lines, showing a slope of $\sim 1.39$ for both $\delta t$ values.}
    \label{fig:trotter_bound}
\end{figure}

Fig.~\ref{fig:lin_fitcross_time} presents a combined analysis of $\tau_c$ across different qubit numbers $n$, which includes data from both Fig.~\ref{fig:fidelity_02} and Fig.~\ref{fig:fidelity_03}. Each curve corresponds to an Erd\H{o}s–R\'enyi graph with varying edge probability $p$, and two Trotter step sizes are considered, $\delta t = 10^{-3}$ and $\delta t = 10^{-2}$. The data are fitted to an exponential curve of the form $T(n) \sim e^{m n + c}$, with the fitted slopes $m$ reported in the legend, $n$ denotes the number of qubits. The results show a clear exponential decay of $\tau_c$ with increasing qubit number. For $\delta t = 10^{-3}$, the fitted slopes vary between $-1.19$ and $-1.27$, with an average value $m_{\text{avg}} \approx {-}\sqrt{3/2}$. For $\delta t = 10^{-2}$, the slopes are slightly steeper, ranging from $-1.32$ to $-1.48$ with an average of $m_{\text{avg}} \approx {-}\sqrt{2}$.

\begin{figure*}[ht!]
    \centering
 \subfloat[ \label{fig:loc_dist_5q_low_p01} $p=0.1$]{\includegraphics[width=\columnwidth]{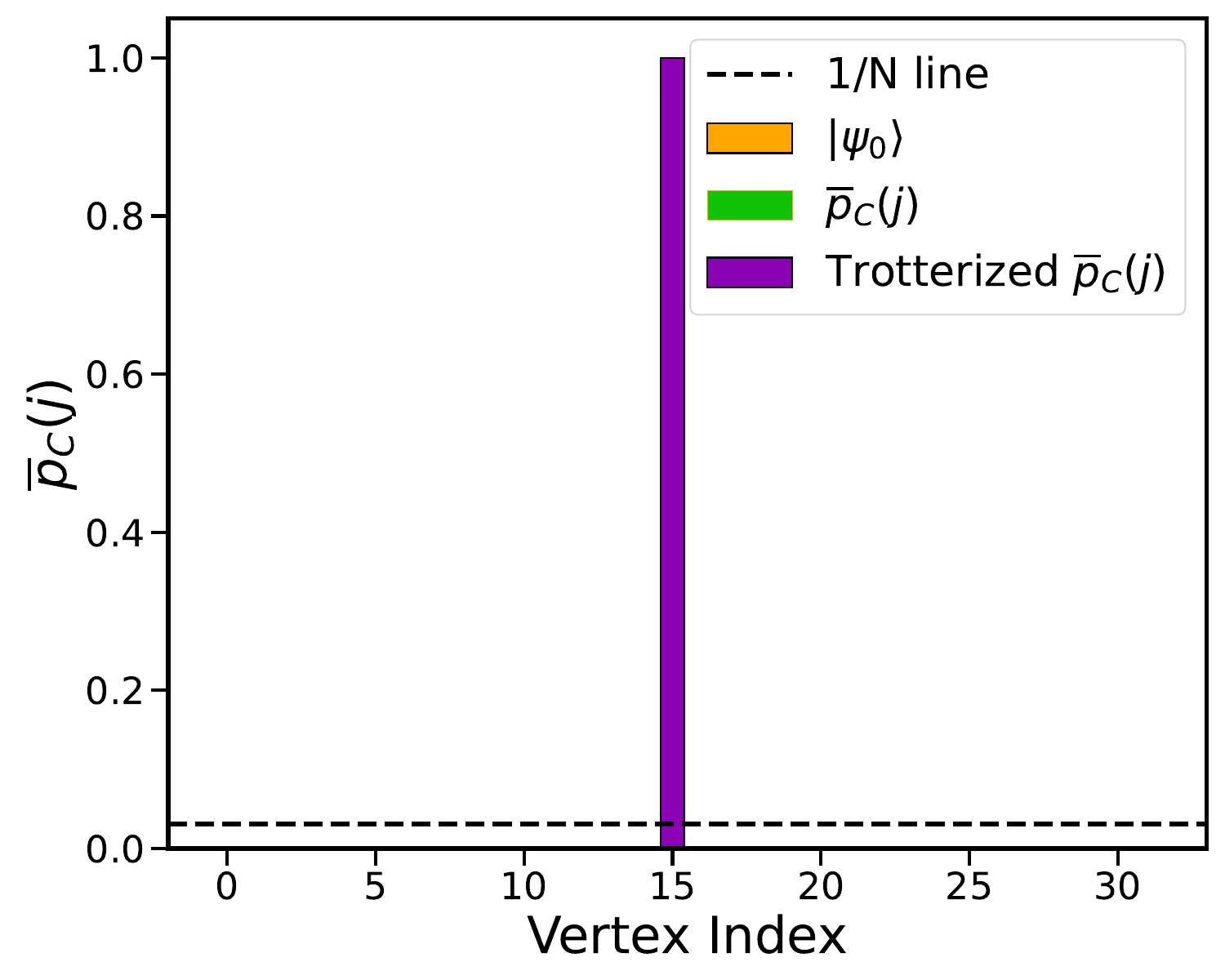}}
 \subfloat[ \label{fig:loc_dist_5q_low_p04} $p=0.4$]{\includegraphics[width=\columnwidth]{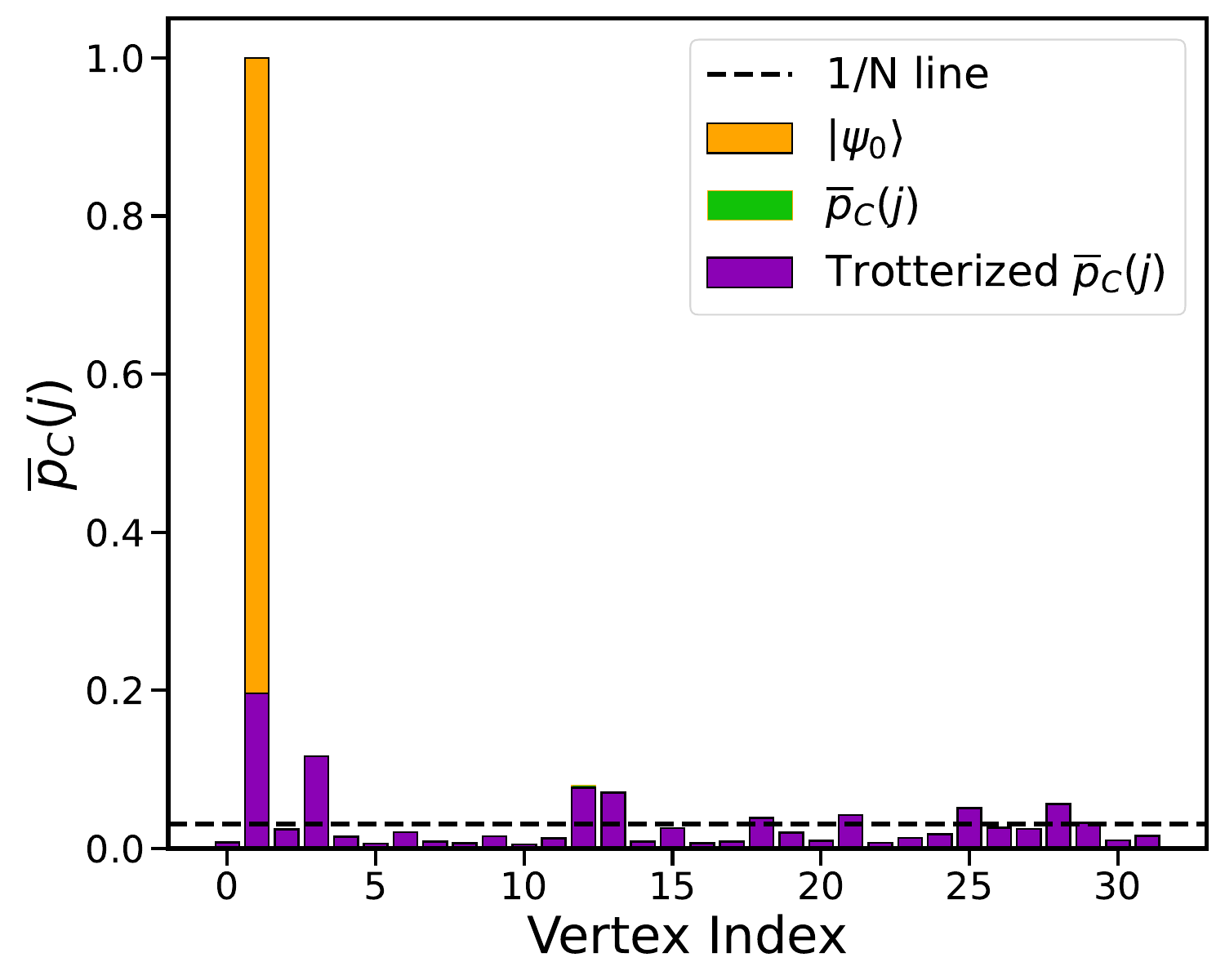}}\\ 
 \subfloat[ \label{fig:loc_dist_cmap_5q_low_p01} $p=0.1$]{\includegraphics[width=\columnwidth]{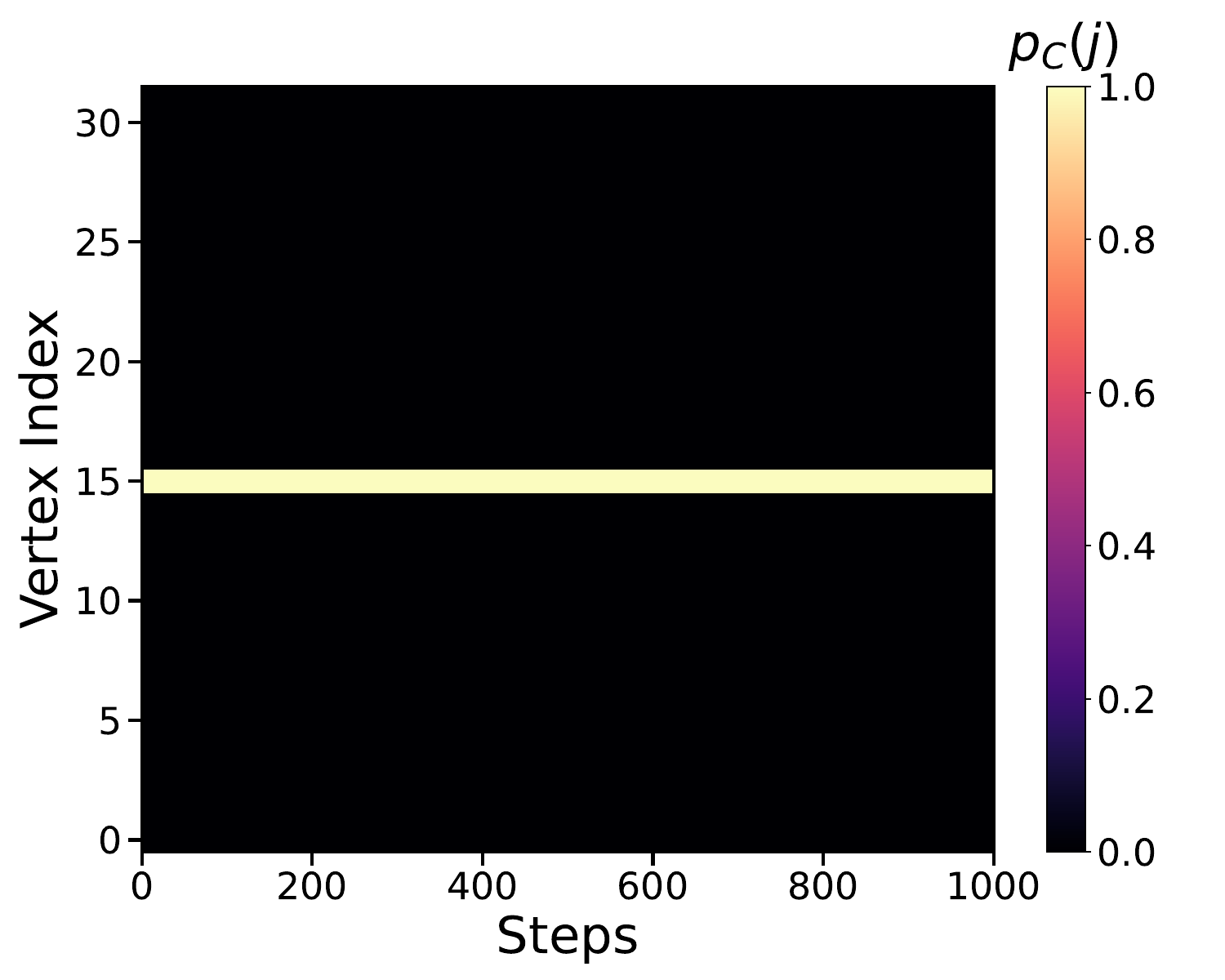}}
 \subfloat[ \label{fig:loc_dist_cmap_5q_low_p04} $p=0.4$]{\includegraphics[width=\columnwidth]{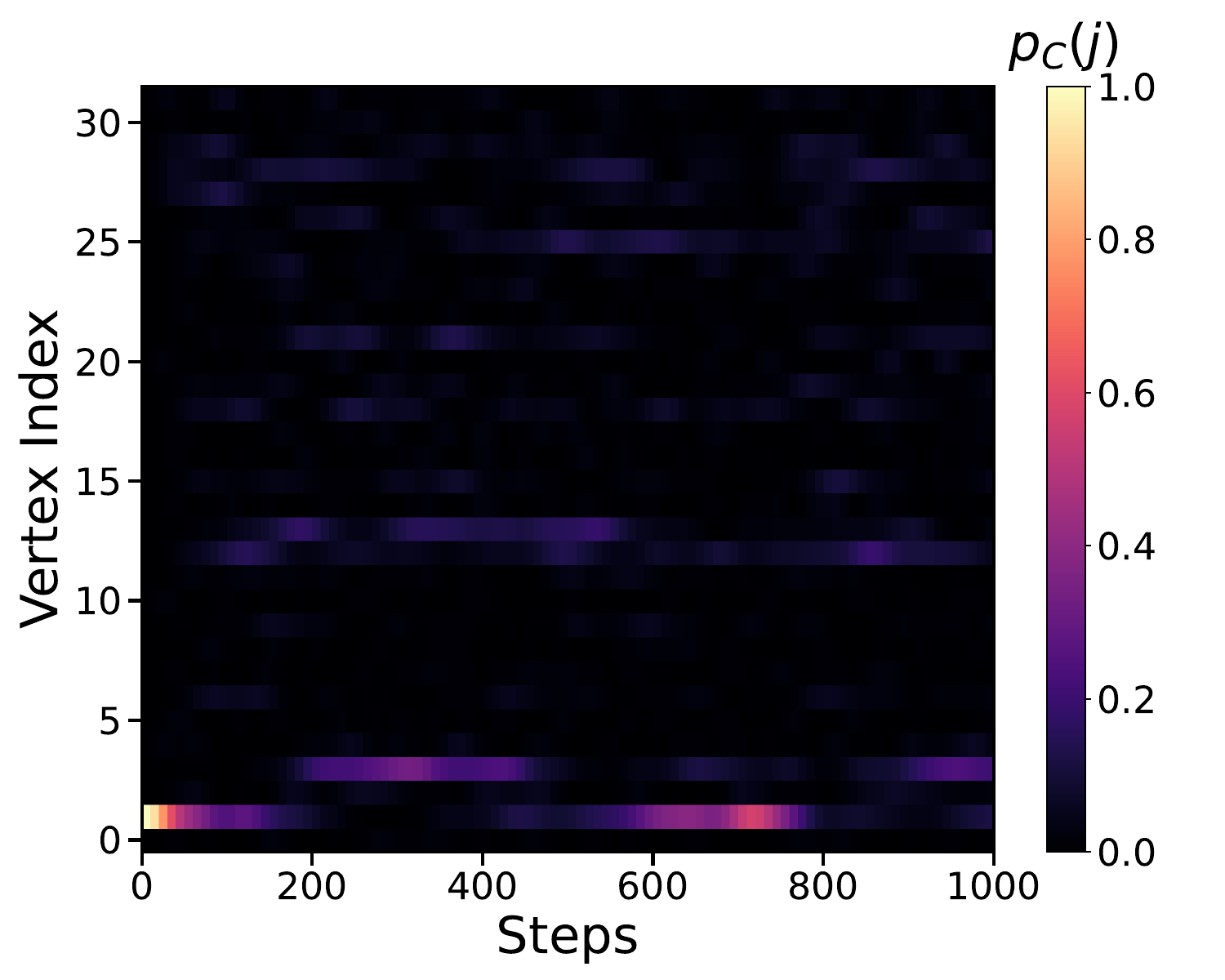}}
    \caption{Panels (a), (b) --- Time-averaged probability distribution $\overline{p}_{c}(j)$ (localization profile) of the quantum walker over all $N = 2^n\:(n=5)$ vertices for different Erd\H{o}s-R\'enyi graph edge probabilities $p = 0.1, 0.4$. The orange bars mark the initial vertex, chosen as the node with the minimum degree. The deviation from the uniform line at $1/N$ indicates varying degrees of localization. Strong peaks at the initial site highlight the persistence of the walker's probability near its origin, even for higher $p$. Results from exact simulation and Trotterized circuit evolution are shown to agree closely. Panels (c), (d) --- Contour plots showing the temporal evolution of the CTQW probability distribution ($p_{c}(j)$) for different edge probabilities $p = 0.1, 0.4$. Initial vertex, chosen as the node with the minimum degree. Each heatmap displays the walker’s probability at each vertex as a function of time. The presence of persistent high-probability bands indicates localization near the initial site. These results are from the circuit-based implementation.}
    \label{fig:loc_dist_5q_low}
\end{figure*}

\begin{figure*}[t!]
    \centering
    \subfloat[ \label{fig:loc_dist_5q_high_p04} $p=0.4$]{\includegraphics[width=0.9\columnwidth]{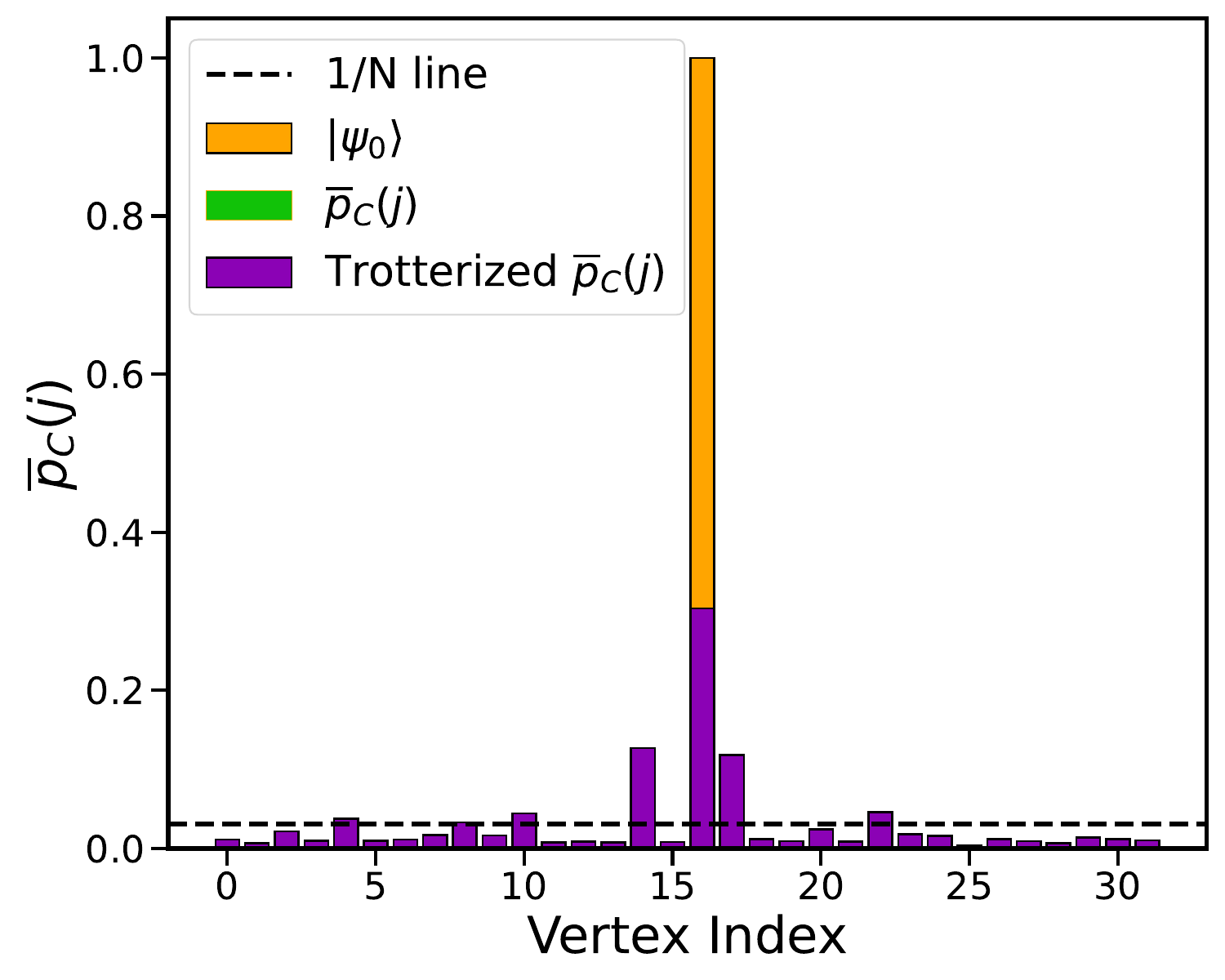}}
    \subfloat[ \label{fig:loc_dist_5q_high_p02} $p=0.7$]{\includegraphics[width=0.9\columnwidth]{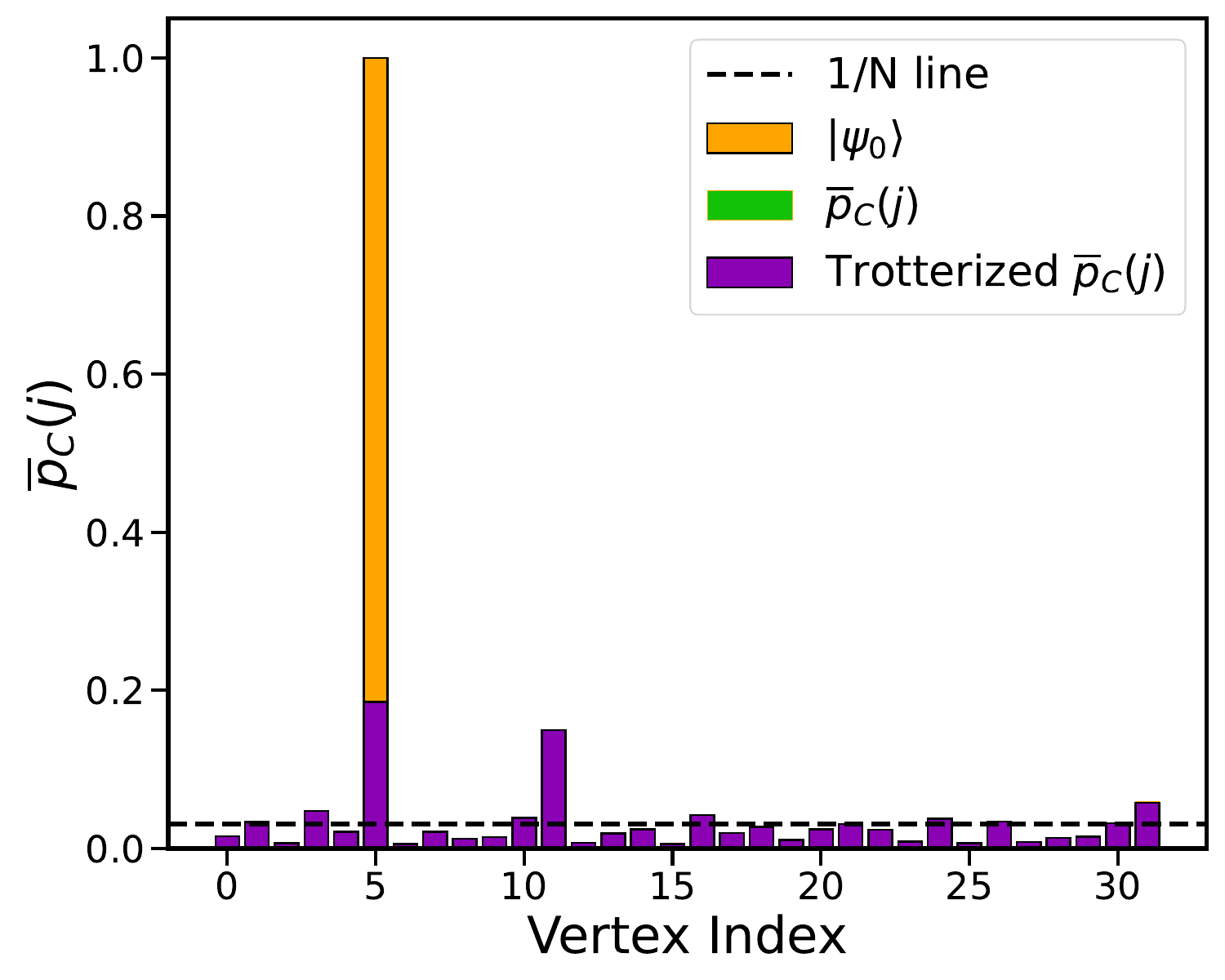}}\\
    \subfloat[ \label{fig:loc_dist_cmap_5q_high_p04} $p=0.4$]{\includegraphics[width=\columnwidth]{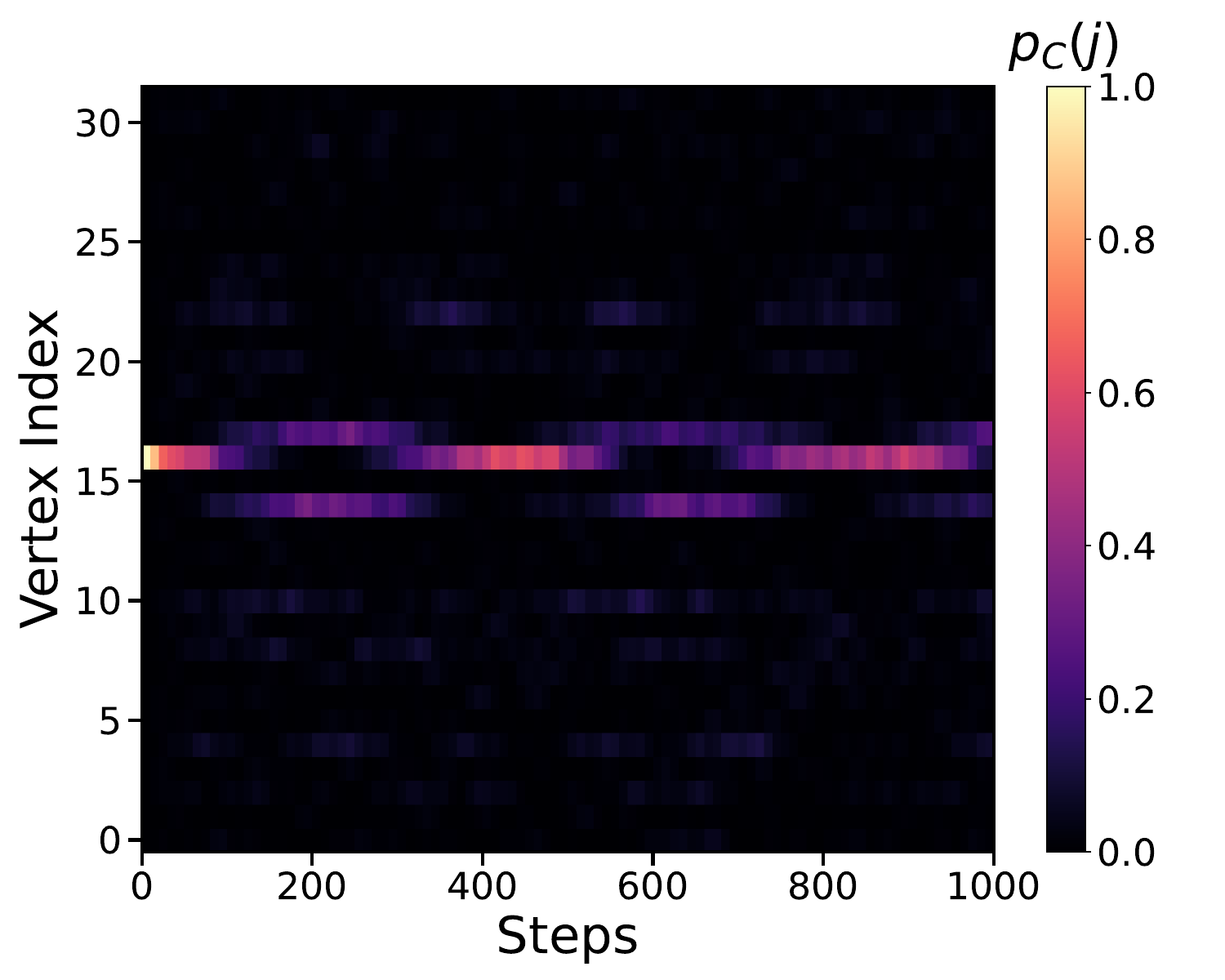}} 
 \subfloat[ \label{fig:loc_dist_cmap_5q_high_p07} $p=0.7$]{\includegraphics[width=\columnwidth]{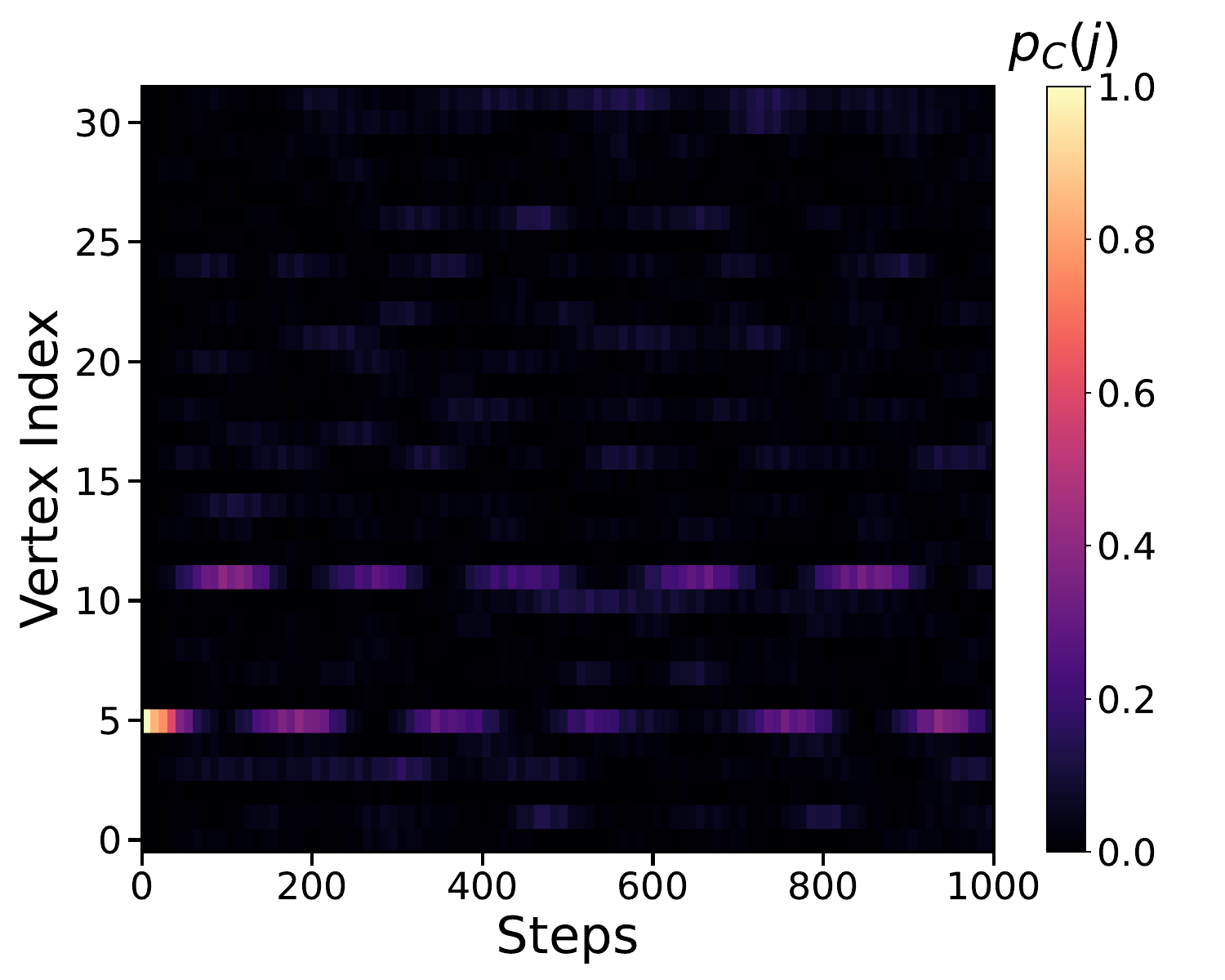}}
 \caption{Panels (a), (b) --- Time-averaged probability distribution $\overline{p}_{c}(j)$ (localization profile) of the quantum walker over all $N = 2^n\:(n=5)$ vertices for different Erd\H{o}s-R\'enyi graph edge probabilities $p = 0.4, 0.7$ at steps 1000. The orange bars mark the initial vertex, chosen as the node with maximum degree. The deviation from the uniform line at $1/N$ indicates varying degrees of localization. Strong peaks at the initial site highlight the persistence of the walker's probability near its origin, even for higher $p$. Results from exact simulation (green bar) and Trotterized circuit evolution (purple bar) are shown to agree closely. Panels (c), (d) --- Contour plots of showing the temporal evolution of the CTQW probability distribution ($p_{c}(j)$) for different edge probabilities $p = 0.4, 0.7$. Initial vertex, chosen as the node with the maximum degree. Each heatmap displays the walker’s probability at each vertex as a function of time. The presence of persistent high-probability bands indicates localization near the initial site. These results are from the circuit-based implementation.}
 \label{fig:loc_dist_5q_high}
\end{figure*}

\begin{figure*}[t!]
    \centering
    \subfloat[ \label{fig:loc_dist_5q_high_p03} $p=1.0$]{\includegraphics[width=0.9\columnwidth]{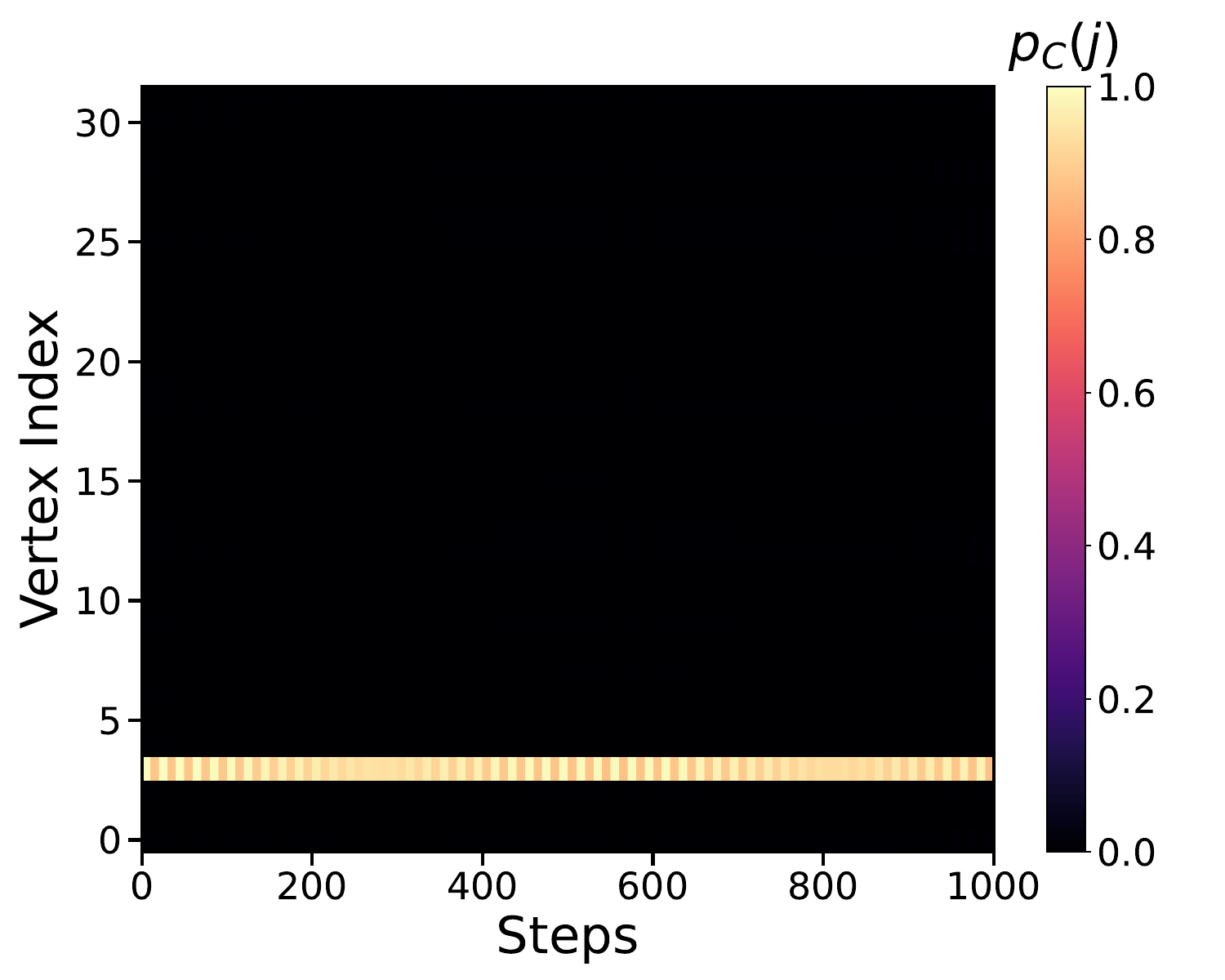}}
    \subfloat[ \label{fig:loc_dist_5q_high_p07} $p=1.0$]{\includegraphics[width=0.9\columnwidth]{trotter_ctqw_contour_p3_high.pdf}}
    \caption{Contour plots of showing the temporal evolution of the CTQW probability distribution ($p_{c}(j)$) for edge probabilities $p = 1.0$. Initial vertex, chosen as the node with maximum degree for $N=2^n$ vertices with (a) $n = 5$ and (b) $n = 6$. Each heatmap displays the walker’s probability at each vertex as a function of time. The presence of persistent high-probability bands indicates localization near the initial site. These results are from the circuit-based implementation.}
    \label{fig:p=1}
\end{figure*}

\begin{figure*}[t!]
    \centering
    \subfloat[ \label{fig:loc_dist_cmap_6q_high_p01} $p=0.1$]{\includegraphics[width=0.9\columnwidth]{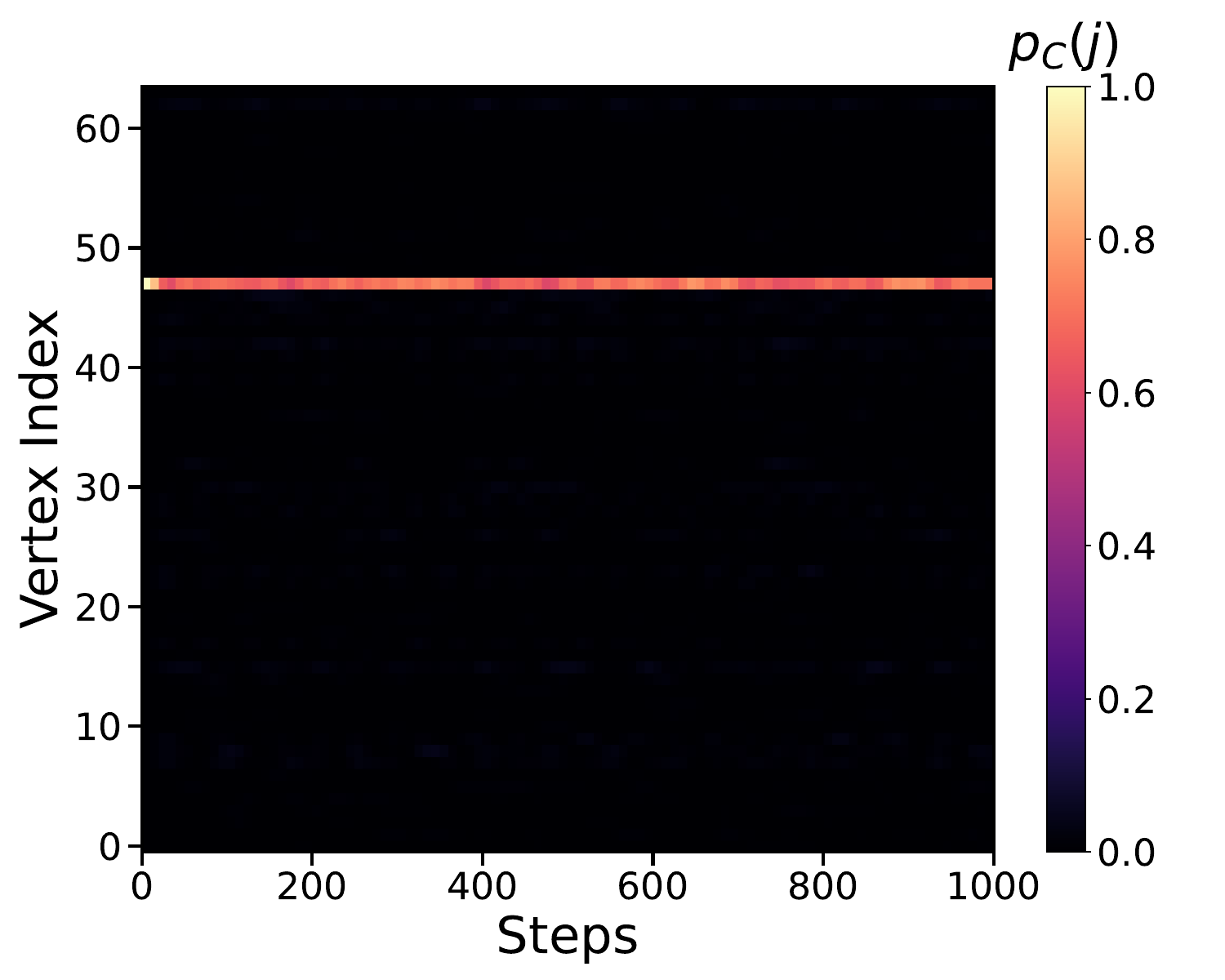}}
    \subfloat[ \label{fig:loc_dist_cmap_6q_high_p02} $p=0.4$]{\includegraphics[width=0.9\columnwidth]{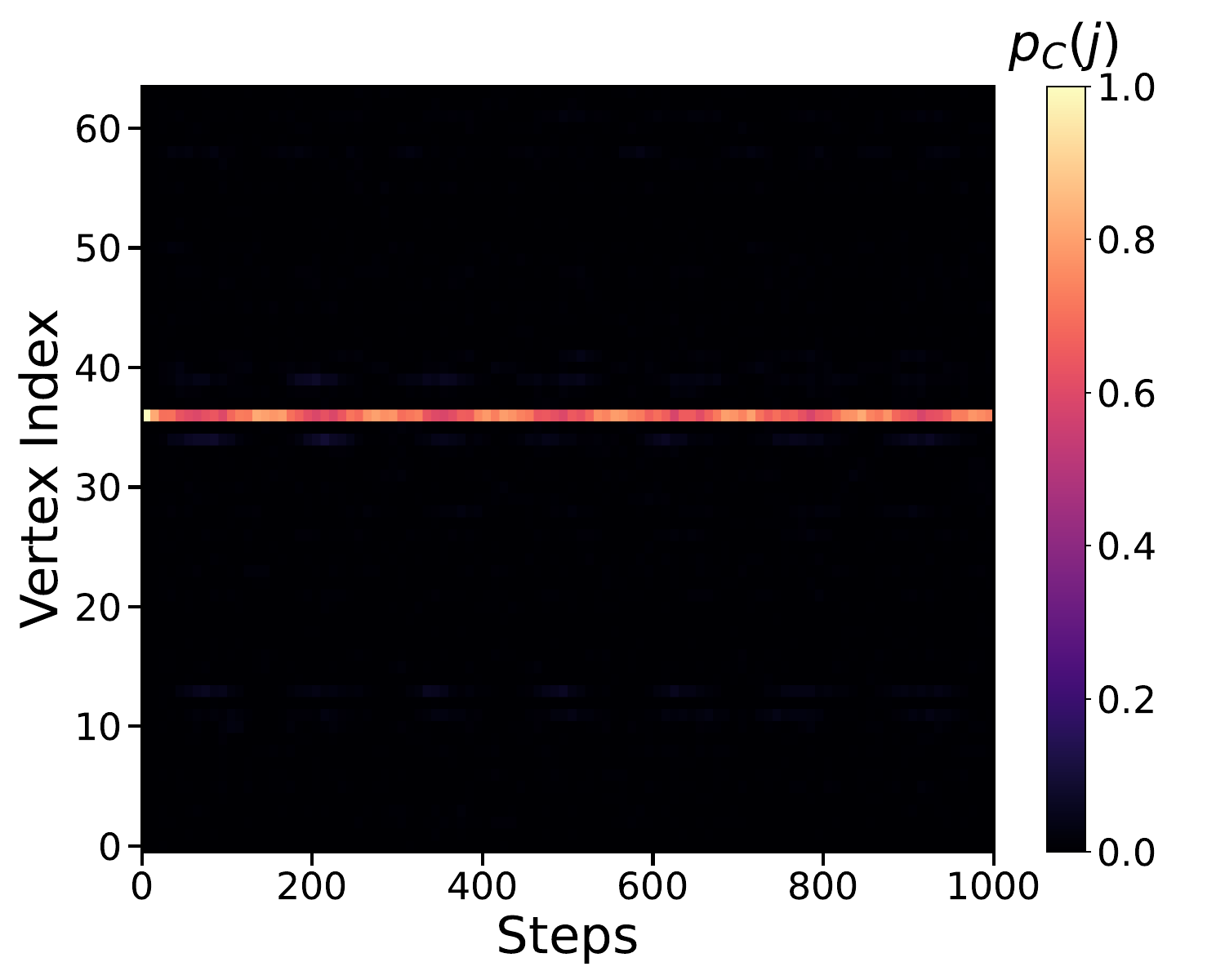}}\\
    \subfloat[ \label{fig:loc_dist_cmap_6q_high_p03} $p=0.7$]{\includegraphics[width=0.9\columnwidth]{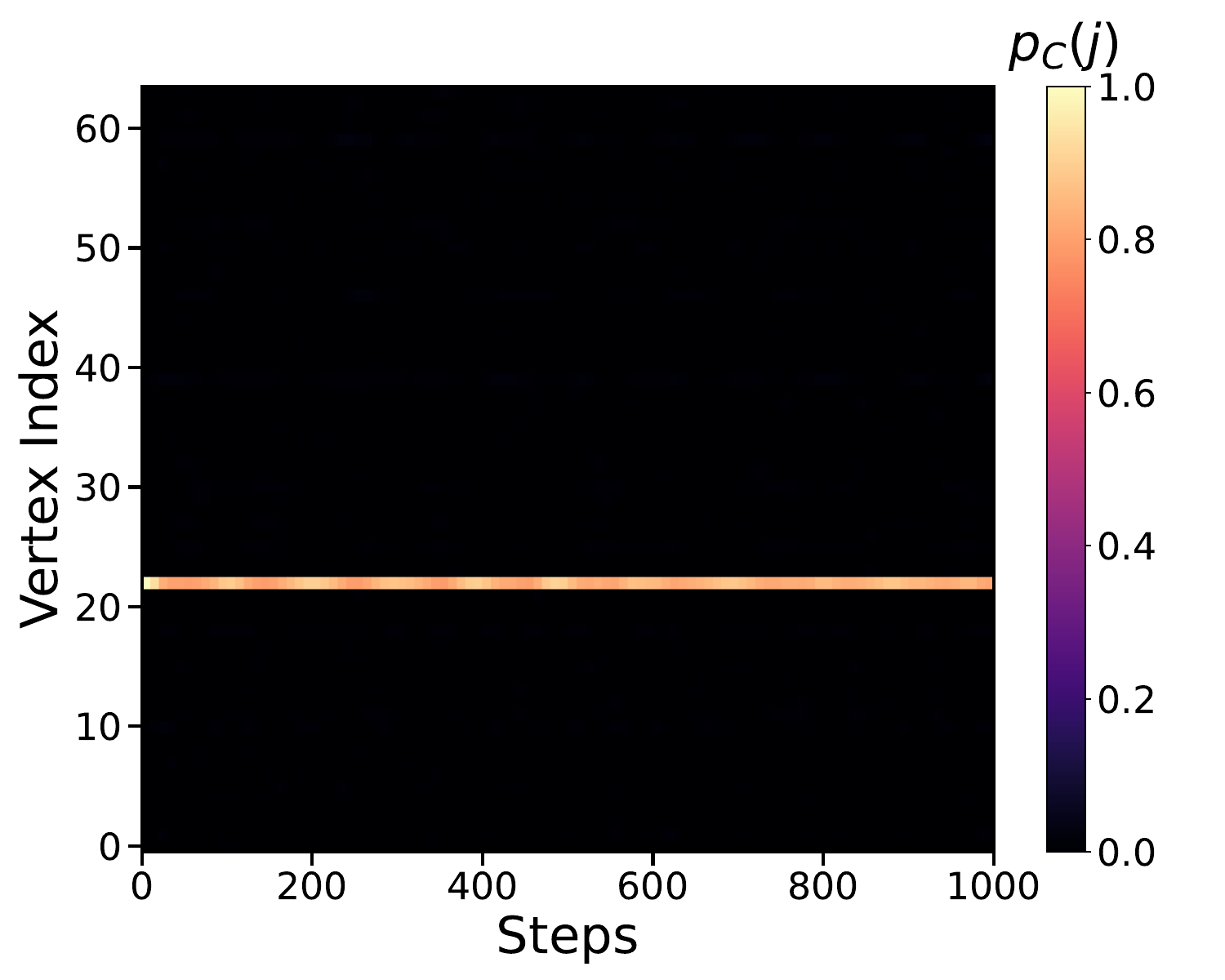}}
    \subfloat[ \label{fig:loc_dist_cmap_6q_high_p04} $p=1.0$]{\includegraphics[width=0.9\columnwidth]{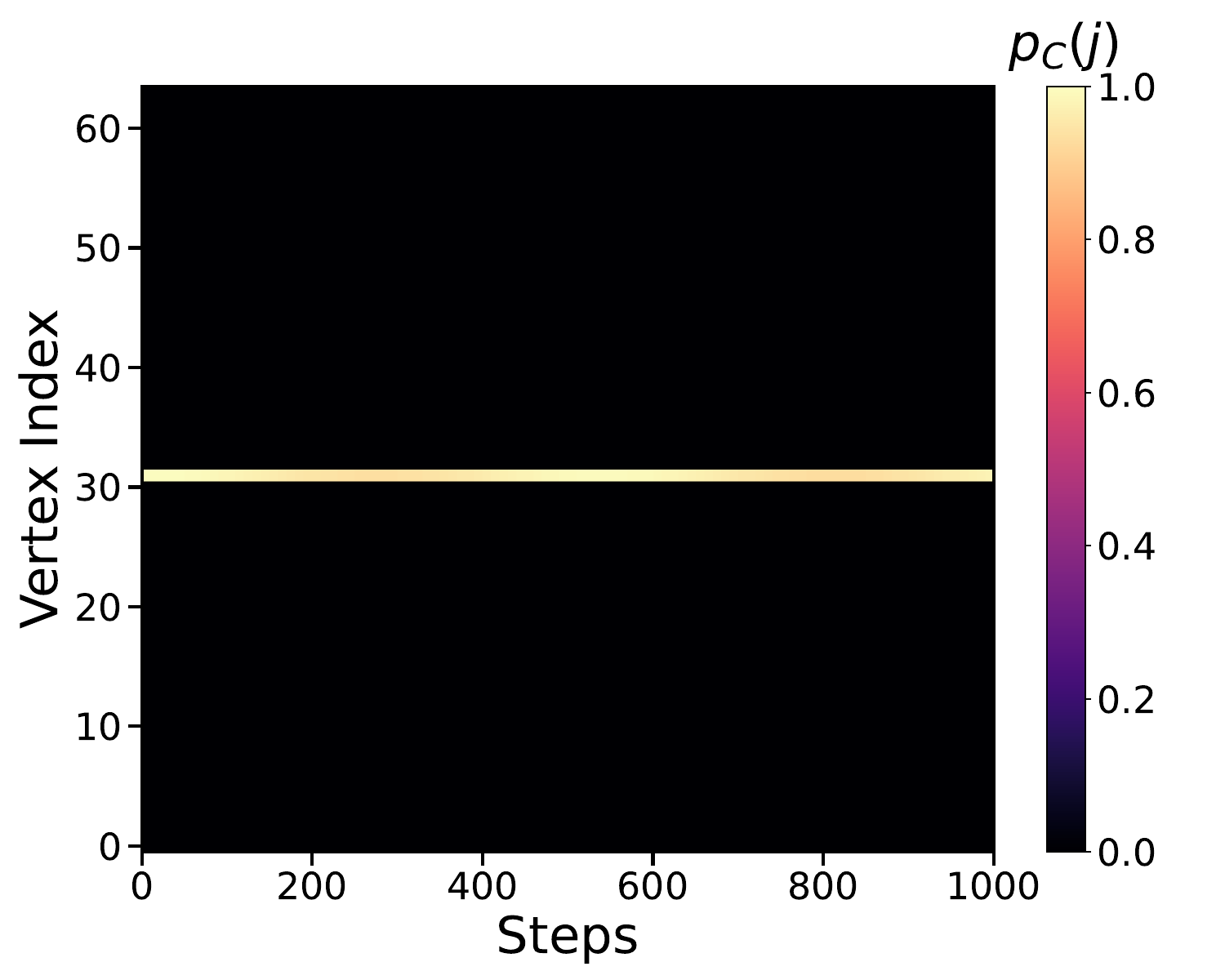}}
    \caption{Panels (a) -- (d) --- Contour plots showing the temporal evolution of the CTQW probability distribution ($p_{c}(j)$) for different edge probabilities $p = 0.1, 0.4, 0.7, 1.0$ (from top left to bottom right) for $N=2^n$ vertices with $n=6$. Initial vertex, chosen as the node with the maximum degree. Each heatmap displays the walker’s probability at each vertex as a function of time. The presence of persistent high-probability bands indicates localization near the initial site. These results are from the circuit-based implementation.}
    \label{fig:loc_dist_cmap_6q_high}
\end{figure*}

To connect these observations with theoretical error estimates, we analyze the scaling of the theoretical Trotter error per step. For a Hamiltonian decomposed into non-commuting terms, the first-order Trotter–Suzuki bound scales as~\cite{childs2021theory}
\begin{equation}
    \varepsilon_{\delta t} \;\sim\; \delta t^2 \, \epsilon \, 2^{2n-1},
\end{equation}
where $\epsilon$ denotes the typical operator norm of commutators among Hamiltonian blocks. In our case, the Laplacian decomposition produces $2^n - 1$ non-commuting blocks, giving rise to approximately $\binom{2^n-1}{2} \sim 2^{2n-1}$ commutator contributions, thereby explaining the exponential scaling of the Trotter error. For a total evolution time $T$ and step size $\delta t$, the accumulated error scales as
\begin{equation}
    \varepsilon_{\text{tot}} \;\sim\; T \cdot \delta t \cdot \epsilon \cdot 2^{2n-1}.
\end{equation}

Fig.~\ref{fig:trotter_bound} shows the scaling of the Trotter error $\varepsilon_{\delta t}$ with $n$, together with theoretical upper bounds, considering $\epsilon=1$. The fitted slope of the error curves is $\sim 1.39$, which is close in magnitude to the average negative slope of the fidelity decay ($m_{\text{avg}} \approx -\sqrt{3/2}$) observed in Fig.~\ref{fig:lin_fitcross_time}. This correspondence indicates that the observed fidelity decay is governed by the exponential growth of Trotter error with qubit number.

It is worth noting that the empirical slopes are somewhat smaller than the theoretical upper bounds. This discrepancy arises because the worst-case analysis overestimates the theoretical error. The effective commutator norms $\epsilon$ are reduced by the sparsity and structure of the Laplacian blocks, and the actual error accumulation depends on the choice of initial state also. These findings validate the effectiveness of the proposed Trotterized circuit architecture for simulating continuous-time quantum walks, while clarifying the limitations imposed by Trotter error scaling.

\section{Localization in CTQW Circuit Simulations}
\label{sec:loc_res}

In this section, we study localization in continuous-time quantum walks. We use localization as a tool for validating the accuracy of the Trotterized circuit evolution against exact simulations. Localization plays a key role in characterizing transport efficiency, memory retention of initial states, and spectral features of the underlying graph Hamiltonian. Unlike Anderson-type localization, which arises from disorder-induced destructive interference, the localization observed here emerges from spectral degeneracies of the graph Hamiltonian~\cite{inui2004localization, chakraborty2020fast, mandal2022limit, bueno2020null, balachandran2024disorder}.

Figs.~\ref{fig:loc_dist_5q_low_p01},~\ref{fig:loc_dist_5q_low_p04},~\ref{fig:loc_dist_5q_high_p04}, and~\ref{fig:loc_dist_5q_high_p02} present the time-averaged probability distributions $\overline{p}_{c}(j)$ of a walker over all $N=2^n$ vertices with $n=5$ for Erd\H{o}s--R\'enyi graphs with edge probabilities $p = 0.4$ and $p = 0.7$ evaluated at $1000$ steps, computed both from exact evolution and from the Trotterized quantum circuit. The orange bars denote the initial vertex ($\ket{\psi_{0}}$), selected as the node with minimum degree for Fig.~\ref{fig:loc_dist_5q_low} and maximum degree for Fig.~\ref{fig:loc_dist_5q_high}. The deviations from the uniform baseline $1/N$ reveal the presence of localization, where we observe a high peak at the initial site ($\ket{\psi_{0}}$), indicating a higher probability of finding the walker near $\ket{\psi_{0}}$. In both cases (exact evolution and the Trotterized circuit evolution), the agreement between the two methods is excellent. A key observation from our simulations is that the degree of the initial vertex strongly influences localization. For Erd\H{o}s–R\'enyi graphs with lower connectivity $p \sim 0.1$, localization becomes particularly pronounced when the walker begins at the vertex of minimum degree.

Apart from it, an interesting observation occurs in the contour plots of the Fig.~\ref{fig:loc_dist_cmap_5q_low_p01},~\ref{fig:loc_dist_cmap_5q_low_p04},~\ref{fig:loc_dist_cmap_5q_high_p04}, and~\ref{fig:loc_dist_cmap_5q_high_p07} where the temporal evolutions of $p_{c}(j)$ are depicted. Each heatmap illustrates the probability distribution across vertices as a function of time. A striking feature emerges for some graphs where vertices that are directly connected and share the same degree show oscillatory behavior in the walker’s probability amplitude when we choose any of them as our initial starting state $\ket{\psi_{0}}$. In such cases, the walker dynamically redistributes its localization weight between these same degree vertices or oscillating vertex group, leading to a persistent oscillation of probability across time. Conversely, other vertices, with the same maximal degree, do not participate in this oscillation if that is not directly connected to the oscillating vertex group. For that vertex, the walker’s localization probability remains comparatively high throughout the evolution if the starting $\ket{\psi_0}$ is on that vertex. This behavior describes the role of graph connectivity. 

In summary, when we initialize the walker at a vertex that carries the maximum degree, the walker tends to localize in that vertex (Figs.~\ref{fig:loc_dist_cmap_6q_high_p01}--\ref{fig:loc_dist_cmap_6q_high_p03}) if it is not directly connected to the oscillating vertex group. This effect originates from the spectral structure of the Laplacian, where high-degree vertices contribute disproportionately to degenerate (or nearly degenerate) eigenmodes. Since the initial state has a large overlap with these modes, part of the amplitude acquires only global phases during evolution, preventing complete delocalization. As a result, the walker retains a significant long-time probability at the starting vertex. Even when $p \geq 0.9$, \emph{i.e.,} when the underlying graph is complete or near-complete, all vertices have the same degree, if we initialize the walker at a single vertex, the time-averaged probability indicates that the walker remains localized at that vertex instead of spreading uniformly across the graph (Figs.~\ref{fig:p=1} and \ref{fig:loc_dist_cmap_6q_high_p04}). This localization does not stem from disorder, as in Anderson localization, but rather from the symmetry and spectral degeneracy~\cite{inui2004localization, chakraborty2020fast, mandal2022limit, bueno2020null, balachandran2024disorder} of the complete graph Laplacian. The decomposition of the initial state into a stationary uniform component and a degenerate oscillatory subspace explains the persistence of amplitude near the origin (a detailed account is given in the Appendix~\ref{appC:loc_p=1}). The complete graph, therefore, provides a striking example where strong connectivity and high symmetry induce localization in CTQWs through purely spectral mechanisms.

\section{Conclusion}
\label{sec:conclusion}

In this work, we have developed a scalable quantum circuit framework for simulating continuous-time quantum walks (CTQWs) on arbitrary random graphs, with a particular focus on Erd\H{o}s--R\'enyi (ER) graphs. By representing the CTQW Hamiltonian in terms of the graph Laplacian and introducing the graph \emph{Laplacian partitioning algorithm} (LPA), we demonstrated that the Laplacian $\bm{L}$ of an $n$-qubit graph can be decomposed into a set of sparse submatrices $\{\bm{L}^{(j)}\}$, each of which is permutation-similar to a block-diagonal form with $2\times2$ non-trivial blocks. This decomposition allows the efficient encoding of the graph Hamiltonian into quantum circuits through permutation matrices that can be realized using \cn~gates.

The resulting framework enables the implementation of the full time-evolution operator $U(t) = e^{-iHt}$ using a Trotter–Suzuki product formula applied to the partitioned Hamiltonian components produced by the LPA. Compared to standard Pauli-string decompositions, where the number of terms scales as $O(4^n)$ with $O(N^2\log N)$ time complexity, our LPA approach achieves a reduced number of terms $O(2^n-1)$ in $O(N^2)$ time. This provides a resource-efficient route for realizing CTQWs on near-term quantum devices and paves the way for the exploration of random graph dynamics on noisy intermediate-scale quantum (NISQ) hardware.

Furthermore, we compared the Trotterized circuit evolution against exact simulations by verifying fidelity of the Trotterized evolution against exact dynamics. The time-averaged probability distributions revealed excellent agreement between exact and circuit-based dynamics, confirming high fidelity of the implemented evolution. We showcase that our circuit error closely follows the theoretical Trotter error. We also tested our circuit using localization as a diagnostic tool. We found that localization in our CTQW implementation arises not from disorder, as in Anderson-type localization, but from spectral degeneracies of the Laplacian. The degree of the initial vertex strongly influences localization strength. The walkers initialized at low-degree vertices in sparse ER graphs ($p\!\sim\!0.1$) exhibit localization, while in dense or complete graphs ($p\!\to\!1$) localization persists due to symmetry-induced degeneracies. In highly connected graphs, oscillatory behavior between connected vertices of equal degree was observed, corresponding to coherent population transfer within degenerate eigen-subspaces. These results demonstrate that spectral structure and graph connectivity dictate localization behavior in CTQWs, and that the proposed circuit framework faithfully reproduces these quantum transport features.

We establish a general framework for Hamiltonian simulation using the graph Laplacian partition algorithm with reduced complexity compared to standard Pauli decomposition. However, we believe that the partitioning strategy could be further improved to have a better fidelity response over larger Trotter steps. This work also opens up the implementation of weighted graph walks, \emph{i.e.,} lackadaisical quantum walks, quantum walks with memories, to name a few. One of the major drawbacks of our method lies in its scalability---as the number of qubits increases, the circuit depth also proportionately increases because of the presence of a higher number of partitions in the LPA. Therefore, optimizing our algorithm to produce fix gate-depth circuit remains a future objective. We can also implement various quantum walk algorithmic tasks, such as the traveling salesman problem~\cite{Marsh_2020_PRR}, finding the inverse of a matrix~\cite{kay_2025_MatrixInversion}. Our work implements CTQW on quantum circuits for random graphs, which is a crucial result at the age of NISQ devices.

\section{Acknowledgment}
SC acknowledge the support of the Prime Minister’s Research Fellowship (PMRF). RKR acknowledges the support by the Institute for Basic Science in Korea (IBS-R024-D1). The authors would also like to acknowledge \emph{Paramshakti} Supercomputer facility at IIT Kharagpur---a national supercomputing mission of the Government of India, for providing the necessary high-performance computational resources.

\appendix
\section{Permutation Operator and Permutation Similarity between Pauli strings}

\subsection{Permutation Operator}
\label{appA:permutation}

The product involving \cn~gates (see Eq.~\eqref{eq:Permut_def}) are equivalent to permutation operation. To understand this argument we follow the work \citet{sarkar2024scalable}. The key idea here is that for a set of $n$-length Pauli strings $\mathcal{S}^{(n)}_P$ ($n$ is the qubit number), one can find a set of permutation matrices $\bm{P}_n^j$, that transform the Pauli strings into block-diagonal matrices with $2\times 2$ non-trivial blocks.  

In Ref.~\cite{sarkar2024scalable} the authors define permutations as $\Pi \mathsf{T}^{e}_{n,x}$ and $\Pi \mathsf{T}^o_{n,x}$\footnote{Here, $\Pi$ denotes the product over all disjoint 2-cycle permutations $P(\alpha,\beta)$ defined by the index functions $\alpha^g_{\kappa_x}$ and $\beta^g_{\kappa_x}$, while $\mathsf{T}$ is simply a symbolic label used to distinguish the corresponding permutation type.}, where $x$ is a binary string
\begin{equation}\label{eqA:x_binary_string}
    x = \sum_{\kappa_j=0}^{n-2} x_{\kappa_j} 2^{\kappa_j}\equiv (x_{n-2},\dots,x_0) \text{ and }x_{\kappa_j} \in \{0,1\}.
\end{equation}

We can also define the index set $\kappa = \set{\kappa_j}$ for each $x$ as $\kappa_x$. These permutations are similar to 

\begin{equation}
\bm{P}_n^j = 
\begin{cases}
    \Pi\mathsf{T}^o_{n,x} = \Pi\mathsf{T}^o_{n,\frac{j}{2}}, & \text{if } j \text{ is even}, \\
    \Pi\mathsf{T}^e_{n,x} = \Pi\mathsf{T}^e_{n,\frac{j-1}{2}}, & \text{otherwise}.
\end{cases}
\end{equation}
Both $x=0$ and $j=0,1$ will give the Identity matrix. 

The notation $e$ ($o$) in $\Pi \mathsf{T}^{e}_{n,x}$ ($\Pi \mathsf{T}^{o}_{n,x}$) indicates that the corresponding permutation matrix is a product of permutations of disjoint $2$-cycles $P(\alpha,\beta)$, where both $\alpha$ and $\beta$\footnote{$\alpha$, $\beta$ are row or column index---which will be clear from the given context.} are even ( exactly one of $\alpha$ or $\beta$ is odd). The notation $P(\alpha,\beta)$ denotes the matrix obtained by exchanging the $\alpha^\th$ and $\beta^\th$ rows of the target matrix.

\begin{proposition}\label{prop:constructeven}
For any $x$, and any $u=(u_{n-2},\dots,u_0)$; define $\bar u_k:=u_k\oplus1$. Consider the functions $\alpha_{\kappa_x}^g:  \{0,1\}^{n-1}\to \set{0,\hdots,2^{n-1}-1}$ and  $\beta_{\kappa_x}^g:  \set{0,1\}^{n-1}\to \{0,\hdots,2^{n-1}-1},$ $g\in\set{e,o}$ defined as

\begin{align*}
\alpha^g_{\kappa_x}(u)
&= \sum_{k\in\kappa_x} u_k\,2^{k+1}
 + \sum_{{\tilde{k}}\notin\kappa_x} u_{\tilde{k}}\,2^{\tilde{k}+1}
 + 2,\\
\beta^e_{\kappa_x}(u)
&= \sum_{k\in\kappa_x} \bar u_k\,2^{k+1}
 + \sum_{{\tilde{k}}\notin\kappa_x} u_{\tilde{k}}\,2^{\tilde{k}+1}
 + 2,\\
\beta^o_{\kappa_x}(u)
&= \sum_{k\in\kappa_x} \bar u_k\,2^{k+1}
 + \sum_{{\tilde{k}}\notin\kappa_x} u_{\tilde{k}}\,2^{\tilde{k}+1}
 + 1.
\end{align*}
Then
\begin{equation*}
\Pi\mathsf T^g_{n,x}
= \prod_{\scalebox{0.8}{$\substack{0\le u<2^{n-1}-1\\ \alpha^g_{\kappa_x}(u)<\beta^g_{\kappa_x}(u)}$}}
P_{\scalebox{0.7}{$\left(\alpha^g_{\kappa_x}(u),\,\beta^g_{\kappa_x}(u)\right)$}},
\quad
g\in\{e,o\}.
\end{equation*}
\end{proposition}

It follows for $x\neq y$, $\Pi\mathsf T^g_{n,x} \neq \Pi\mathsf T^g_{n,y}$ with
\begin{align*}
\left(\alpha^g_{\kappa_x}(u),\,\beta^g_{\kappa_x}(u)\right)
&\neq
\left(\alpha^g_{\kappa_y}(u),\,\beta^g_{\kappa_y}(u)\right)\\
&\text{for all } 0 \le u \le 2^{n-1}-1.
\end{align*}

Example: for $n=3$ and $x=1$, from Eq.~\eqref{eqA:x_binary_string}---

\begin{equation*}
x=1 \implies (x_1,x_0)=(0,1)\implies\quad \kappa_x=\{0\}.
\end{equation*}

\begin{align*}
    \begin{split}
        &\alpha_{\kappa_x}^e(u)=2u_0+4u_1+2,\\
        &\beta_{\kappa_x}^e(u)=2(1-u_0)+4u_1+2,\\
        &u=(u_1,u_0)\in\{0,1\}^2.
    \end{split}
\end{align*}

\begin{equation*}
\begin{array}{c|cc|c}
u & \alpha_{\kappa_x}^e(u) & \beta_{\kappa_x}^e(u) & \alpha<\beta\\
\hline
(0,0) & 2 & 4 & \text{yes}\\
(0,1) & 4 & 2 & \text{no}\\
(1,0) & 6 & 8 & \text{yes}\\
(1,1) & 8 & 6 & \text{no}
\end{array}
\end{equation*}
We have,
\begin{align*}
&\Pi\mathsf T^e_{3,1}\\
&=
\prod_{\scalebox{0.8}{$\substack{u\in\{0,1\}^2\\ \alpha_{\kappa_x}^e(u)<\beta_{\kappa_x}^e(u)}$}}
P_{\scalebox{0.8}{$(\alpha_{\kappa_x}^e(u),\,\beta_{\kappa_x}^e(u))$}}
=
P_{(2,4)}\,P_{(6,8)}.
\end{align*}
Similarly for 
\begin{align*}
&\Pi\mathsf T^o_{3,1}\\
&=
\prod_{\scalebox{0.8}{$\substack{u\in\{0,1\}^2\\ \alpha_{\kappa_x}^o(u)<\beta_{\kappa_x}^o(u)}$}}
P_{\scalebox{0.8}{$(\alpha_{\kappa_x}^o(u),\,\beta_{\kappa_x}^o(u))$}}
=
P_{(2,3)}\,P_{(6,7)}.
\end{align*}

\begin{align*}
P^{j=2}_{n=3}= \Pi\mathsf T^o_{3,1} &:=& ~ \vcenter{\hbox{\Qcircuit @C=1em @R=1em {
\lstick{1} & \qw   & \qw         & \qw      & \qw\\
\lstick{2} & \ctrl{1}        & \targ           & \ctrl{1}           & \qw\\
\lstick{3} & \targ      & \ctrl{-1}     & \targ         & \qw\\
} }}\\
\vspace{2pt}\\
P^{j=3}_{n=3} = \Pi\mathsf T^e_{3,1} &:= &~\vcenter{\hbox{\Qcircuit @C=1em @R=1em {
\lstick{1} & \qw     & \qw \\
\lstick{2} & \targ       & \qw \\
\lstick{3} & \ctrl{-1}  & \qw \\
} }}
\end{align*}

\subsection{Permutation Similarity between Pauli strings}
\label{appA:permutation_sim_Pauli}

We now state the following theorem from Ref.~\cite{sarkar2024scalable}, which establishes permutation similarity of the elements of $\mathcal{S}_{I, X}^{(n)}$. 
\begin{theorem}\label{th:SRBBvPauli}\cite{sarkar2024scalable}
Let $j\in \{1,\hdots,2^{n}-1\}.$ Then  
\begin{align*}
    \begin{split}
        \mathcal{S}_{\{I,X\}_{j}}^{(n)}&=\begin{cases}
    \Pi\mathsf{T}_{n,\frac{j-1}{2}}^e (I_2^{\otimes (n-1)}\otimes X) \Pi\mathsf{T}_{n,\frac{j-1}{2}}^e,~j \text{ odd}, \\
    \Pi\mathsf{T}_{n,\frac{j}{2}}^o (I_2^{\otimes (n-1)}\otimes X) \Pi\mathsf{T}_{n,\frac{j}{2}}^o,~j \text{ even},
\end{cases}
    \end{split}\\
& = \bm{P}_n^j (I_2^{\otimes (n-1)}\otimes X) \bm{P}_n^j.
\end{align*}
\end{theorem}

\begin{proof} 
See \citet{sarkar2024scalable} for the detailed proof. 
\end{proof}

\section{Lemma 3 Example}\label{appB:Lemma_3}

Consider the circuit~\ref{circ:ZYZ_decomp3} with two controls (top wires) and one target (bottom wire). From left to right, between the single-qubit rotations on the target, the \cn s from the controls to the target are connected. We denote the four target rotations by $R_z(\omega_1), R_z(\omega_2), R_z(\omega_3), R_z(\omega_4)$.

Here $k=2$, so $c\in\{00,01,10,11\}$ is the control basis string. Just before each rotation, the active-control mask is

\begin{equation}
    m_1=00,\; m_2=01,\; m_3=10,\; m_4=11.
\end{equation}

Interpret each mask $m_i$ via its \emph{overlap size} \emph{i.e.} (number of shared 1s). The parity used in Lemma~\ref{lem:angle_parity} is precisely this overlap size mod 2.

Hence, the unitary is block diagonal,
\begin{equation}
    U=\bigoplus_{c\in\{0,1\}^2} R_z(\eta_c),
\; \text{ where }
\eta_c=\sum_{i=1}^4(-1)^{\,\langle c\cdot m_i \rangle}\,\omega_i.
\end{equation}

For each control string $c$ we list the overlap sizes $\langle c\cdot m_i \rangle =: o_{i}$ (with $i=1,\ldots,4$), their parities, and the resulting signs:

\begin{widetext}
    \begin{equation}
    \begin{array}{c|cccc|cccc|c}
c & o_1 & o_2 & o_3 & o_4 & o_1\!\!\!\!\!\!\pmod 2 & o_2\!\!\!\!\!\!\pmod 2 & o_3\!\!\!\!\!\!\pmod 2 & o_4\!\!\!\!\!\!\pmod 2 & \text{signs }((-1)^{o_i})\\
\hline
00 & 0 & 0 & 0 & 0 & 0 & 0 & 0 & 0 & (+,+,+,+)\\
01 & 0 & 1 & 0 & 1 & 0 & 1 & 0 & 1 & (+,-,+,-)\\
10 & 0 & 0 & 1 & 1 & 0 & 0 & 1 & 1 & (+,+,-,-)\\
11 & 0 & 1 & 1 & 2 & 0 & 1 & 1 & 0 & (+,-,-,+)
\end{array}
\end{equation}
\end{widetext}

Using the signs above in $\eta_c=\sum_i (-1)^{o_i(c)}\omega_i$ gives
\begin{align}
\begin{split}
\eta_{00} &= \omega_1+\omega_2+\omega_3+\omega_4,\\
\eta_{01} &= \omega_1-\omega_2+\omega_3-\omega_4,\\
\eta_{10} &= \omega_1+\omega_2-\omega_3-\omega_4,\\
\eta_{11} &= \omega_1-\omega_2-\omega_3+\omega_4.
\end{split}
\end{align}

Equivalently, with $\omega=(\omega_1,\omega_2,\omega_3,\omega_4)^{T}$,
\begin{equation}\label{eqB:transformation}
    \mqty(
\eta_{00}\\ \eta_{01}\\ \eta_{10}\\ \eta_{11}
)
=
\underbrace{\mqty(
+1&+1&+1&+1\\
+1&-1&+1&-1\\
+1&+1&-1&-1\\
+1&-1&-1&+1
)}_{2^{2}H^{\otimes 2}}
\mqty(
\omega_1\\ \omega_2\\ \omega_3\\ \omega_4
).
\end{equation}

\section{Localization profile for all connected graphs}
\label{appC:loc_p=1}

To quantify localization, we employ the inverse participation ratio (IPR), defined for a walker initialized at vertex $j$ as, 

\begin{equation}\label{eqC:IPR}
    \mathrm{IPR}_j(t) = \sum_{i=1}^N p_{ij}^2(t), ~~ 
    p_{ij}(t) = \left| \bra{i} e^{-iHt} \ket{j} \right|^2 ,
\end{equation}
which measures the spread of the probability distribution in the vertex basis. For a completely delocalized state, $p_{ij}(t) \approx 1/N$ for all vertices, yielding $\mathrm{IPR}_j(t) \approx 1/N$, which serves as a natural ergodic baseline. Localization is implied at a said vertex $j$, whenever the value of IPR at that vertex is greater than $1/N$. 

For a complete graph $K_N$, each vertex is connected to all others with degree 
\begin{equation}
    \deg(v) = N - 1 \qquad \text{for all } v \in K_N.
\end{equation}

The CTQW Hamiltonian is defined as (assuming $\gamma=1$)
\begin{equation}
    H = -\bm{L},
\end{equation}
where $\bm{L}$ is the Laplacian of $K_N$. The evolution operator is $U(t) = \exp(i \bm{L} t)$, and its spectral decomposition governs the transport dynamics. The Laplacian spectrum of the complete graph is highly degenerate: there is one eigenvalue $E_1 = 0$, corresponding to the uniform superposition state, and $(N-1)$ degenerate eigenvalues equal to $N$,
\begin{equation}
    E_1 = 0, \qquad E_j = N \quad (j = 2, \dots, N).
\end{equation}
This large degeneracy underpins the persistence of localization in CTQWs on $K_N$.

The normalized eigenvector associated with the zero eigenvalue is the uniform state
\begin{equation}\label{eqC:expansion}
    \ket{s} = \frac{1}{\sqrt{N}} \sum_{j=1}^{N} \ket{j},
\end{equation}
while the remaining eigenvectors span the subspace orthogonal to $\ket{s}$. An initial state localized at a vertex $\ket{v}$ can be decomposed as
\begin{equation}\label{eqC:decomposition}
    \ket{v} = \braket{s}{v} \ket{s} + \Big( \ket{v} - \braket{s}{v}\ket{s} \Big).
\end{equation}
The uniform component $\ket{s}$ is stationary since $L\ket{s}=0$, whereas the orthogonal component evolves with a global phase $e^{iNt}$, owing to its eigenvalue $N$. The total state at time $t$ is therefore
\begin{equation}
    \ket{\psi(t)} = \braket{s}{v}\,\ket{s} + e^{iNt}\Big(\ket{v}-\braket{s}{v}\ket{s}\Big).
\end{equation}

The amplitude to remain at the initial vertex is
\begin{equation}
    \braket{v}{\psi(t)} = \frac{1}{N} + \left(1-\frac{1}{N}\right)e^{iNt},
\end{equation}
leading to the instantaneous probability
\begin{align}
\begin{split}
    &\left|\braket{v}{\psi(t)}\right|^2 \\
    &= \frac{1}{N^2} + \left(1-\frac{1}{N}\right)^2 
    + \frac{2}{N}\left(1-\frac{1}{N}\right)\cos(Nt).
\end{split}
\end{align}
Averaging over time removes the oscillatory term and yields the time-averaged probability at the starting vertex,
\begin{equation}
    \overline{p}_v = \frac{1}{N^2} + \left(1-\frac{1}{N}\right)^2 = 1 - \frac{2}{N} + \frac{2}{N^2}.
\end{equation}
For large $N$, this approaches
\begin{align}
    \overline{p}_v &\approx 1 - \frac{2}{N},
\end{align}
which is markedly higher than the uniform distribution $1/N$. Thus, even though the complete graph is maximally connected, the walker retains a strong probability of being detected at its initial position at long times.

To compute the IPR, we first note that using Eq.~\eqref{eqC:expansion}
\begin{equation}
\ip{i}{\psi(t)} = \frac{1}{N} + e^{iNt}\left(\delta_{iv}- \frac{1}{N}\right).    
\end{equation}
Which further allows us to write from Eq.~\eqref{eqC:decomposition}
\begin{align}
    \begin{split}
        p_{iv}(t) & = \abs{\frac{1}{N} + \left(1-\frac{1}{N}\right)e^{iNt}}^2 ~\text{for }i=v,\\
        &=  1-\frac{2}{N} + \frac{2}{N^2} + \frac{2(N-1)}{N^2}\cos{Nt}.
    \end{split}\\
    \begin{split}
    p_{iv}(t) & = \abs{\frac{1}{N}(1 - e^{iNt})}^2 ~ \text{for } i\neq v, \\
    & =  \frac{2}{N^2}(1 - \cos{Nt}).
    \end{split}
\end{align}
Therefore, we can compute the IPR using Eq.~\eqref{eqC:IPR} as
\begin{align}
    &\mathrm{IPR}_v(t) \notag\\
\begin{split}
    &= \underbrace{p^2_{iv}(t)}_{i=v} + (N-1)\underbrace{p^2_{iv}(t)}_{i\neq v},\\
    &= 1 - \frac{4}{N}+ \frac{10}{N^2} - \frac{6}{N^3} + \frac{4(N-1)(N-2)}{N^3}\cos{Nt},\\
    &+ \frac{2(N-1)}{N^3}\cos{2Nt}.
\end{split}
\end{align}
The average over a long time results in 
\begin{equation}
    \overline{\mathrm{IPR}}_v = 1 - \frac{4}{N}+ \frac{10}{N^2} - \frac{6}{N^3},
\end{equation}
which in the large $N$ limit reduces to 
\begin{equation}
    \overline{\mathrm{IPR}}_v \approx 1 - \frac{4}{N}.
\end{equation}
This suggests that for large $N$, on average, the IPR remains close to 1, suggesting strong localization.

\bibliography{ctqw_qckt.bib}
\end{document}